\documentclass[a4paper,fleqn,usenatbib]{mnras}

\usepackage{newtxtext,newtxmath}
\usepackage{comment}

\usepackage[T1]{fontenc}
\usepackage{ae,aecompl}

\usepackage{graphicx}	
\usepackage{amsmath}	
\usepackage{amssymb}	
\usepackage{multirow}
\usepackage{subfig}
\usepackage{fixltx2e}

\usepackage{color}
\usepackage[dvipsnames]{xcolor}

\newcommand{\herschel}{\textit{Herschel}}
\newcommand{\lambdar}{\textsc{lambdar}}
\newcommand{\magphys}{\textsc{magphys}}

\newcommand{\pVmaxalpha}{-1.22 \pm 0.01}
\newcommand{\pVmaxConalpha}{-1.22 \pm 0.01}
\newcommand{\pVmaxTmpalpha}{-1.21 \pm 0.01}
\newcommand{\bbdSBalpha}{-1.24 \pm 0.02}

\newcommand{\bbdDMalpha}{-1.27 \pm 0.01}
\newcommand{\bbdDMConalpha}{-1.27 \pm 0.01}

\newcommand{\totlowalpha}{-1.24 \pm 0.02}

\newcommand{\pVmaxMstar}{4.65 \pm 0.18}
\newcommand{\pVmaxConMstar}{4.39 \pm 0.17}
\newcommand{\pVmaxTmpMstar}{4.47 \pm 0.16}
\newcommand{\bbdSBMstar}{4.32 \pm 0.17}

\newcommand{\bbdDMMstar}{4.67 \pm 0.15}
\newcommand{\bbdDMConMstar}{4.40 \pm 0.15}

\newcommand{\totlowMstar}{3.72 \pm 0.15}

\newcommand{\pVmaxphistar}{6.26 \pm 0.28}
\newcommand{\pVmaxConphistar}{6.49 \pm 0.30}
\newcommand{\pVmaxTmpphistar}{6.97 \pm 0.30}
\newcommand{\bbdSBphistar}{6.94 \pm 0.36}

\newcommand{\bbdDMphistar}{5.65 \pm 0.23}
\newcommand{\bbdDMConphistar}{5.85 \pm 0.24}

\newcommand{\totlowphistar}{6.36 \pm 0.45}

\newcommand{\pVmaxOmegad}{1.11 \pm 0.02}
\newcommand{\pVmaxConOmegad}{1.08 \pm 0.02}
\newcommand{\pVmaxTmpOmegad}{1.17 \pm 0.02}
\newcommand{\bbdSBOmegad}{1.11 \pm 0.02}

\newcommand{\bbdDMOmegad}{1.11 \pm 0.02}
\newcommand{\bbdDMConOmegad}{1.08 \pm 0.02}

\newcommand{\totlowOmegad}{0.92 \pm 0.02}

\newcommand{\pVmaxOmegadinf}{1.11 \pm 0.02}

\newcommand{\bbdDMOmegadinf}{1.06 \pm 0.01}

\newcommand{\MoffElowalpha}{-1.01 \pm 0.08}
\newcommand{\MoffnotElowalpha}{-1.18 \pm 0.03}

\newcommand{\MoffElowMstar}{0.98 \pm 0.23}
\newcommand{\MoffnotElowMstar}{4.17 \pm 0.25}

\newcommand{\MoffElowphistar}{2.05 \pm 0.41}
\newcommand{\MoffnotElowphistar}{5.75 \pm 0.48}

\newcommand{\MoffElowOmegad}{0.060 \pm 0.005}
\newcommand{\MoffnotElowOmegad}{0.88 \pm 0.03}

\newcommand{\MoffElowOmegadinf}{0.060 \pm 0.005}
\newcommand{\MoffnotElowOmegadinf}{0.88 \pm 0.03}

\newcommand{\MoffElowOmegadsum}{0.060 \pm 0.005}

\newcommand{\pVmaxDSFalphaone}{-1.29 \pm 0.08}

\newcommand{\bbdDMDSFalphaone}{-1.33 \pm 0.15}
\newcommand{\pVmaxConDSFalphaone}{-1.29 \pm 0.13}

\newcommand{\bbdDMConDSFalphaone}{-1.33 \pm 0.17}
\newcommand{\pVmaxDSFalphatwo}{1.85 \pm 1.69}

\newcommand{\bbdDMDSFalphatwo}{2.07 \pm 1.69}
\newcommand{\pVmaxConDSFalphatwo}{2.32 \pm 1.68}

\newcommand{\bbdDMConDSFalphatwo}{2.42 \pm 1.86}
\newcommand{\pVmaxDSFMstarone}{4.65 \pm 0.55}

\newcommand{\bbdDMDSFMstarone}{4.59 \pm 0.73}
\newcommand{\pVmaxConDSFMstarone}{4.16 \pm 0.95}

\newcommand{\bbdDMConDSFMstarone}{4.05 \pm 1.20}
\newcommand{\pVmaxDSFMstartwo}{0.89 \pm 0.44}

\newcommand{\bbdDMDSFMstartwo}{0.75 \pm 0.43}
\newcommand{\pVmaxConDSFMstartwo}{0.78 \pm 0.45}

\newcommand{\bbdDMConDSFMstartwo}{0.71 \pm 0.54}
\newcommand{\pVmaxDSFphistar}{6.15 \pm 2.72}

\newcommand{\bbdDMDSFphistar}{5.47 \pm 5.80}
\newcommand{\pVmaxConDSFphistar}{6.04 \pm 5.28}

\newcommand{\bbdDMConDSFphistar}{5.61 \pm 8.97}
\newcommand{\pVmaxDSFfmix}{0.80 \pm 0.17}

\newcommand{\bbdDMDSFfmix}{0.81 \pm 0.17}
\newcommand{\pVmaxConDSFfmix}{0.85 \pm 0.18}

\newcommand{\bbdDMConDSFfmix}{0.85 \pm 0.18}
\newcommand{\pVmaxDSFOmegad}{1.11 \pm 0.02}

\newcommand{\bbdDMDSFOmegad}{1.11 \pm 0.02}
\newcommand{\pVmaxConDSFOmegad}{1.11 \pm 0.02}

\newcommand{\Dunnealpha}{-1.01^{+0.17}_{-0.14}}
\newcommand{\DunneMstar}{3.9^{+0.74}_{-0.63}}
\newcommand{\Dunnephistar}{8.09^{+1.9}_{-1.72}}
\newcommand{\DunneOmegad}{1.01 \pm 0.15}
\newcommand{\Vlakalpha}{-1.39^{+0.03}_{-0.02}}
\newcommand{\VlakMstar}{6.0^{+0.45}_{-0.55}}
\newcommand{\Vlakphistar}{3.33^{+0.74}_{-0.5}}
\newcommand{\VlakOmegad}{0.94 \pm 0.44}
\newcommand{\Clemalpha}{-1.34 \pm 0.4}
\newcommand{\ClemMstar}{5.27 \pm 1.56}
\newcommand{\Clemphistar}{4.78 \pm 1.81}
\newcommand{\ClemOmegad}{1.1 \pm 0.22}
\newcommand{\MstETGBnEMt}{0.94^{+0.25}_{-0.24}}
\newcommand{\rhoETGBnEMt}{1.00\pm0.11}

\newcommand{\MstLTGBndskMt}{7.77^{+0.76}_{-0.73}}
\newcommand{\rhoLTGBndskMt}{10.21\pm0.45}
\newcommand{\MsttotBndskMt}{6.93^{+0.61}_{-0.58}}
\newcommand{\rhototBndskMt}{10.66\pm0.38}
\newcommand{\rhoLTGBnLTGMt}{8.07\pm0.35}

\title[A census of dust in galaxies to $z=0.1$]{GAMA/{\bf \it H-}ATLAS: The Local Dust Mass Function and Cosmic Density as a Function of Galaxy Type - A Benchmark for Models of Galaxy Evolution}

\author[R. A.~Beeston et al.]{R. A.~Beeston$^{1}$\thanks{E-mail: BeestonRA@cardiff.ac.uk (RAB)},
A. H.~Wright$^{2,3}$,
S.~Maddox$^{1,4}$,
H.L.~Gomez$^{1}$,
L.~Dunne$^{1,4}$, \and
S. P.~Driver$^{2}$,
A.~Robotham$^{2}$,
C. J. R.~Clark$^{1}$,
K.~Vinsen$^{2}$,
T. T.~Takeuchi$^{5}$,  \and
G.~Popping$^{6,7}$,
N.~Bourne$^{8}$,
M. N.~Bremer$^{9}$,
S.~Phillipps$^{9}$,
A. J.~Moffett$^{2}$, \and
M.~Baes$^{10}$,
J. Bland-Hawthorn$^{11}$,
S. Brough$^{12}$,
P. De Vis$^{13}$,
S. A. Eales$^{1}$, \and
B. W. Holwerda$^{14}$,
J. Loveday$^{15}$,
J. Liske$^{16}$,
M. W. L.~Smith$^{1}$, \and
D. J. B.~Smith$^{17}$,
E. Valiante${^1}$,
C. Vlahakis$^{18}$,
L. Wang$^{19,20}$
\\
$^{1}$School of Physics \& Astronomy, Cardiff University, Queens Buildings, The Parade, Cardiff , CF24 3AA, UK\\
$^{2}$ICRAR, The University of Western Australia, 35 Stirling Highway, WA 6009, Australia \\
$^{3}$Arglander-Institut f{\"u}r Astronomie, Universit{\"a}t Bonn, Auf dem H{\"u}gel 71, 53121 Bonn, Germany \\
$^{4}$SUPA. Institute for Astronomy, University of Edinburgh, Royal Observatory, Blackford Hill, Edinburgh, EH9 3HJ, UK \\
$^{5}$Division of Particle and Astrophysical Science, Nagoya University, Furo-Cho, Chikusa-ku, Nagoya 464-8602, Japan\\
$^{6}$European Southern Observatory, Karl-Schwarzschild-Strasse 2, 85748, Garching, Germany\\
$^{7}$Max-Planck-Institut f\"{u}r Astronomie, K\"{o}nigstuhl 17, D-69117 Heidelberg, Germany\\
$^{8}$Institute for Astronomy, University of Edinburgh, Royal Observatory, Edinburgh EH9 3HJ, UK \\
$^{9}$H. H. Wills Physics Laboratory, University of Bristol, Tyndall Avenue, Bristol BS8 1TL, UK\\
$^{10}$Sterrenkundig Observatorium, Universiteit Gent, Krijgslaan 281 S9, B-9000 Gent, Belgium\\
$^{11}$Sydney Institute for Astronomy, School of Physics, A28, The University of Sydney, NSW 2006, Australia \\
$^{12}$ School of Physics, University of New South Wales, NSW 2052, Australia \\
$^{13}$Institut d'Astrophysique Spatiale, CNRS, Universit\'{e} Paris-Sud, Universit\'{e} Paris-Saclay, B\^{a}t. 121, 91405, Orsay Cedex, France\\
$^{14}$Department of Physics and Astronomy, 102 Natural Science Building, University of Louisville, Louisville KY 40292, USA \\
$^{15}$Astronomy Centre, University of Sussex, Falmer, Brighton BN1 9QH, UK\\
$^{16}$Hamburger Sternwarte, Universit\"{a}t  Hamburg,  Gojenbergsweg  112, D-21029 Hamburg, Germany \\
$^{17}$Centre for Astrophyics Research, School of Physics, Astronomy and Mathematics, University of Hertfordshire, College Lane, \\ Hatfield AL10 9AB, UK\\
$^{18}$National Radio Astronomy Observatory, 520 Edgemont Road, Charlottesville, VA 22903, USA \\
$^{19}$SRON Netherlands Institute for Space Research, Landleven 12, 9747 AD, Groningen, The Netherlands\\
$^{20}$Kapteyn Astronomical Institute, University of Groningen, Postbus 800, 9700 AV, Groningen, The Netherlands \\
}

\date{Accepted 2018 May 31. Received 2018 May 31; in original form 2017 December 12}

\pubyear{2018}

\begin{document}
\label{firstpage}
\pagerange{\pageref{firstpage}--\pageref{lastpage}}
\maketitle

\begin{abstract}

  We present the dust mass function (DMF) of 15,750 galaxies with redshift $z< 0.1$, drawn from the overlapping area of the GAMA and {\it H-}ATLAS surveys.  The DMF is derived using the density  corrected $V_{\rm max}$ method, where we estimate $V_{\rm max}$  using: (i) the normal photometric selection limit ($pV_{\rm max}$) and (ii) a bivariate brightness distribution (BBD) technique, which accounts for two selection effects. We fit the data with a Schechter function, and find $M^{*}=(\pVmaxMstar)\times 10^{7}\,h^2_{70}\, M_{\odot}$, $\alpha=(\pVmaxalpha)$, $\phi^{*}=(\pVmaxphistar)\times 10^{-3}\,h^3_{70}\,\rm Mpc^{-3}\,dex^{-1}$.  The resulting dust mass density parameter integrated down to $10^4\,M_{\odot}$ is $\Omega_{\rm d}=(\pVmaxOmegad)\times 10^{-6}$ which implies the mass fraction of baryons in dust is $f_{m_b}=(2.40\pm 0.04)\times 10^{-5}$; cosmic variance adds an extra 7-17\,per\,cent uncertainty to the quoted statistical errors. Our measurements have fewer galaxies with high dust mass than predicted by semi-analytic models. This is because the models include too much dust in high stellar mass galaxies. Conversely, our measurements find more galaxies with high dust mass than predicted by hydrodynamical cosmological simulations. This is likely to be from the long timescales  for grain growth assumed in the models. We calculate DMFs split by galaxy type and find dust mass densities of $\Omega_{\rm d}=(\MoffnotElowOmegadinf)\times 10^{-6}$ and $\Omega_{\rm d}=(\MoffElowOmegadinf)\times 10^{-6}$ for late-types and early-types respectively.  Comparing to the equivalent galaxy stellar mass functions (GSMF) we find that the DMF for late-types is well matched by the GMSF scaled by $(\rhoLTGBnLTGMt) \times 10^{-4}$.

\end{abstract}




\section{Introduction}

Cosmic dust is a significant, albeit small, component of the interstellar medium (ISM) of galaxies.  Despite being less than 1\% of the baryonic mass of a galaxy, dust is responsible for obscuring the ultraviolet and optical light from stars and active galactic nuclei and is thought to have absorbed approximately half of the starlight emitted since the Big Bang \citep{Puget1996,Fixsen1998,Dole2006,Driver2016}.  Measuring the dust mass in galaxies is therefore important for understanding obscured star formation \citep{Kennicutt1998,Calzetti2007,Marchetti2016}, particularly at different cosmic epochs \citep{Madau1998,Hopkins2004,Takeuchi2005}. The dust mass function (DMF) is one of the fundamental measurements of the dust content of galaxies, providing crucial information on the reservoir of metals that are locked up in dust grains \citep{Issa1990,Edmunds2001,Dunne2003}. A measure of the space density of dusty galaxies is becoming even more relevant given the widespread use of dust emission as a tracer for the gas in recent years (\citealt{Eales2010,Eales2012,Magdis2012,Scoville2014,Scoville2017}; see also the comprehensive review of \citealt{Casey2014}). This is of particular interest given difficulties in observing atomic and molecular-line gas mass tracers out to higher redshifts \citep{Tacconi2013,Catinella2015}.

 Ground-based studies including observations at 450 and 850\,$\mu$m with the Submillimetre Common User Bolometer Array (SCUBA) on the James Clerk Maxwell Telescope, led to the first measurements of the DMF over the mass range $ \sim 10^7 {\, M_{\odot}} < M_{\rm d}  < {\rm few}\times 10^8 {\, M_{\odot}}$ \citep{Dunne2000,Dunne2001,Vlahakis2005}, where $M_{\rm d}$ is dust mass. Unfortunately the state-of-the-art at that time meant fewer than 200 nearby galaxies were observed with small fields of view and selected at optical or infrared (60\,$\mu$m) wavelengths.  At higher redshifts, the Balloon-borne Large Aperture Submillimeter Telescope (BLAST, observing at 250-500\,$\mu$m) enabled a DMF to be derived out to $z=1$ \citep{Eales2009} and a valiant effort to measure at even higher redshifts $(z=2.5)$ using SCUBA surveys was attempted by \citet{Dunne2003}. These studies were hampered by small number statistics and difficulties with observing from the ground.

The advent of the \textit{Herschel Space Observatory} (hereafter \herschel, \cite{Pilbratt2010}) and \textit{Planck} Satellite revolutionised studies of dust in galaxies, as they enabled greater statistics, better sensitivity and angular resolution in some regimes, wider wavelength coverage and the ability to observe orders of magnitude larger areas of the sky than possible before.
The largest dust mass function of galaxies using \herschel\, was presented in \citet{Dunne2011} consisting of 1867 sources out to redshift $z=0.5$, selected from the Science Demonstration Phase (SDP) of the \herschel~Astrophysical Terahertz Large Area Survey ({\it H-}ATLAS) blind 250-$\mu$m fields (\citealp{Eales2010}, 16\,sq.\,degrees). Their DMF extended down to $5\times 10^5\, M_{\odot}$ and they derived a redshift dependent dust mass density of $\Omega_{\rm d}=\rho_{\rm d} / \rho_{\rm crit} = (0.7-2)\times 10^{-6}$.  Subsequently, \citet{Negrello2013,Clemens2013} published the DMF of 234 local star-forming galaxies from the all sky \textit{Planck} catalogue. \citet{Clark2015} then derived a local DMF from a 250-$\mu$m selected sample consisting of 42 sources. These DMFs ranged from $10^6 {\,  M_{\odot}} < M_{\rm d} < {\rm few}\times 10^8 {\, M_{\odot}}$ and $2 \times 10^5 {\, M_{\odot}} < M_{\rm d} < 10^8 {\, M_{\odot}}$ respectively. These measurements\footnote{scaled to the same dust absorption coefficient, $\kappa$} were found to be consistent with the $z=0$ estimate from \citet{Dunne2011}, once scaled to the same dust properties, as well as those derived from optical obscuration studies using the Milleniuum Galaxy Catalogue (\citealp{Driver2007}).

Interestingly, although the dust mass density is broadly consistent across most surveys, the shape of the dust mass function differs between all of these different estimates.  \citet{Clark2015} demonstrated using a blind survey selected at 250-$\mu$m, around a third of the dust mass in the local universe is contained within galaxies that are low stellar mass, gas-rich and have very blue optical colours.  These galaxies were shown to have colder dust populations on average ($1 2< T_{\rm d} < 16\,\rm K$, where $T_{\rm d}$ is the cold-component dust temperature) compared to other \herschel\ studies of nearby galaxies, e.g. the \herschel\ Reference Survey \citep{Boselli2010}, the Dwarf Galaxy Survey (\citealp{Madden2013,Remy-Ruyer2013}, see also \citealt{DeVis2017a}) and higher stellar mass {\it H-}ATLAS galaxies \cite{Smith2012}.  This led to higher numbers of galaxies in the low dust mass regime than predicted from extrapolating the \citet{Dunne2011} DMF down to the equivalent mass bins \citep{Clark2015}.

In comparison, the \citet{Clemens2013} and \citet{Vlahakis2005} DMFs are in reasonable agreement and both suggest a low-mass slope that is much steeper than the \citet{Dunne2011} function.  Overall, comparing between these different measures is complex due to different selection effects; furthermore they are limited due to (i) small number statistics, and/or (ii) lack of sky coverage or volume, inflating uncertainties due to cosmic variance. { We also show evidence in Section \ref{sec:dmf} that fitting the same dataset over different mass ranges can have a significant effect on the resulting best-fit parameters. Since we probe further down the low-mass end than any literature study, this could therefore have a significant impact. }

Here we further the study of the DMF by deriving the `local' ($z<0.1$) dust mass function for the largest sample of galaxies to date, the sample is taken from the Galaxy and Mass Assembly Catalogue (GAMA, \citealt{Driver2011}).  The large size of this sample reduces the statistical uncertainties and the effect of cosmic variance. We also employ statistical techniques to address selection effects in our sample, which allows us to probe further down the dust mass function by at least an order of magnitude compared to previous works.  We present the observations and sample selection in Section~\ref{sec:observations} and the method used to derive the dust masses for the GAMA sources in Section~\ref{sec:magphys}. The dust mass function is presented in Section~\ref{sec:dmf} and is compared to predictions from semi-analytical models in Section~\ref{sec:theory}. In Sections~\ref{sec:split_DMF} and Section~\ref{sec:GSMF_comp}, we split the DMF by morphological type and compare with their corresponding stellar mass functions, with conclusions in Section~\ref{sec:conc}.  Properties of the full GAMA sample are discussed in detail in \citet{Driver2017} and the accompanying stellar mass function of the same sample is published in \cite{Wright2017}, hereafter W17. Throughout this work we use a cosmology of $\Omega_m=0.3$, $\Omega_{\Lambda}=0.7$ and $H_0 = 70\,\rm km\,s^{-1}\,Mpc^{-1}$.

\section{Observations and Photometry}
\label{sec:observations}

\subsection{GAMA}

The GAMA\footnote{http://www.gama-survey.org/} survey is a panchromatic compilation of galaxies built upon a highly complete magnitude limited spectroscopic survey of around { 286 square degrees of sky (with limiting magnitude $r_{\rm petro} \leq 19.8\thinspace$mag as measured by the Sloan Digital Sky Survey (SDSS) DR7, \citealp{Abazajian2009}). Around 238,000 objects have been successfully observed with the AAOmega Spectrograph on the Anglo-Australian Telescope as part of the GAMA survey.} As well as spectrographic observations, GAMA has collated broad-band photometric measurements in up to 21 filters for each source from ultraviolet (UV) to far-infrared (FIR)/submillimetre (submm) \citep{Driver2016,Wright2017}. The imaging data required to derive photometric measurements come from the compilation of many other surveys: GALEX Medium Imaging Survey \citep{Bianchi1999}; the SDSS DR7 \citep{Abazajian2009}, the VST Kilo-degree Survey (VST KiDS, \citealp{deJong2013}); the VIsta Kilo- degree INfrared Galaxy survey (VIKING, \citealp{deJong2013}); the Wide-field Infrared Survey Explorer (WISE, \citealp{Wright2010}); and  the {\it Herschel-}ATLAS \citep{Eales2010}. The motivation and science case for GAMA is detailed in \cite{Driver2009}. The GAMA input catalogue definition is described in \cite{Baldry2010}, and the tiling algorithm in \cite{Robotham2010}. The data reduction and spectroscopic analysis can be found in \cite{Hopkins2013b}. An overview and the survey procedures for { the first data release (DR1)} are presented in \citet{Driver2011}. The second data release {(DR2)} was nearly twice the size of the first and is described in \cite{Liske2015}. { Information on data release 3 (DR3) can now be found in \cite{Baldry2018}}. There is now a vast wealth of data products available for the GAMA survey, making it an incredibly powerful database for all kinds of extragalactic astronomy and cosmology.

K-corrections for GAMA sources are available from \cite{Loveday2012} using \textsc{\small k-correct} v4\_2 \citep{Blanton2007}. Redshifts derived using {\sc autoz} are available from \cite{Baldry2014}. This work consists of data from the GAMA equatorial fields, which has a redshift completeness of $>$98\,per\,cent at $r_{\rm petro} \leq 19.8\thinspace$mag \citep{Liske2015}. GAMA distances were calculated using spectroscopic redshifts and corrected \citep{Baldry2012} to account for bulk deviations from the Hubble flow \citep{Tonry2000}.

{ For this paper, we select galaxies in the three equatorial fields of the GAMA survey, which cover $\sim 180$ square degrees of sky between them. { The equatorial fields G09, G12, G15 are located on the celestial equator at roughly 9$\,$h, 12$\,$h, and 15$\,$h, respectively}.  We use the redshift range $0.002 \le z \le 0.1$, with the upper limit matching the low $z$ bin from the earlier DMF study of \citet{Dunne2011}; this redshift range contains 20,387 galaxies (with spectroscopic redshift quality set at $\rm n_{Qual}\ge 3$)\footnote{ Here we use the following GAMA catalogues: LambdarCatv01, SersicCatSDSSv09, VisualMorphologyv03, DistancesFramesv14, and TilingCatv46 and the {\sc magphys} results presented in \citep{Driver2017}. We also removed one galaxy, GAMA CATAID 49167, due to an error in the $r$-band aperture chosen to derive the photometry of this source.}.} These GAMA galaxies have been further split into Early Types (ETGs), Late Types (LTGs) and little blue spheroids (LBSs) based on classifications using $giH$-band images from SDSS \citep{York2000}, VIKING \citep{Sutherland2015} or UKIDSS-LAS (see \citealt{Kelvin2014,Moffett2016_gals} for more details on the classification). 

\subsection{{\it Herschel-}ATLAS}

The FIR and submm imaging data, which are necessary to derive dust masses, are provided via {\it H-}ATLAS\footnote{http://www.h-atlas.org/} \citep{Eales2010}, the largest extragalactic Open Time survey using \herschel. This survey spans {\raise.17ex\hbox{$\scriptstyle\sim$}}660 square degrees of sky and consists of over 600 hours of observations in parallel mode across five bands (100 and 160\,$\mu$m with PACS - \citealt{Poglitsch2010}, and 250, 350, and 500\,$\mu$m with SPIRE - \citealt{Griffin2010}). {\it H-}ATLAS was specifically designed to overlap with other large area surveys such as SDSS and GAMA. The GAMA/{\it H-}ATLAS overlap covers around 145\,sq.\,degrees over the three equatorial GAMA fields, G09, G12, and G15.  Photometry in the five bands for the {\it H-}ATLAS DR1 is provided in \cite{Valiante2016} based on sources selected initially at 250\,$\mu$m using \textsc{madx} (Maddox et al. \textit{in prep.}) and having $S/N > 4$ in any of the three SPIRE bands. \citet{Bourne2016} present optical counterparts to the {\it H-}ATLAS sources, identified from the GAMA catalogue using a likelihood ratio technique \citep{Smith2011}.  In this paper, we use the aperture-matched photometry from \herschel\ based on the GAMA $r$-band aperture definitions using the LAMBDAR package
(\citealp{Wright2016}), this method is described briefly in Section~\ref{sec:lambdar}.

{ Given the requirement for {\it H-}ATLAS and GAMA coverage, the} final sample for this work consists of 15,951 galaxies, this number includes a selection on $r_{\rm petro} \leq 19.8$  and the fact that due to the shapes of the {\it H-}ATLAS and GAMA fields, some of the GAMA sources were not covered by \herschel.

\subsection{Photometry with LAMBDAR}
\label{sec:lambdar}

The Lambda Adaptive Multi-band Deblending Algorithm in R (\lambdar)\footnote{\lambdar~is available from \newline https://github.com/AngusWright/LAMBDAR} is an aperture photometry package developed by \cite{Wright2016}, which performs photometry based on an input catalogue of sources. Aperture-matched photometry can be implemented on any number of bands and for each band the apertures are convolved by the PSF of the instrument. \lambdar~also deblends sources occupying the same on-sky area, this is achieved by sharing the flux in each pixel between all overlapping apertures. The fractional splitting is done iteratively and, depending on user preference, can be based on the mean surface brightness of a source, central pixel flux, or a user-defined weighting system. Each source is considered in a postage stamp of the input image focused on the source, the size of which depends upon the size of the aperture itself. All known sources within the postage stamp are deblended, including an optional list of known contaminants specified by the user. For this paper this includes {\it H-}ATLAS detected sources from \cite{Valiante2016} which do not have a reliable optical counterpart. These are assumed to be higher redshift background sources.

The sky estimate for each source is calculated by randomly placing blank apertures with dimensions equal to the object aperture on the postage stamp, using the number of masked pixels in each blank aperture to weight its contribution to the background estimate. Furthermore, during flux iteration, if any component of a blend is assigned a negative flux then it is rejected for all subsequent iterations (and any negative measurement is set to zero). There are a very small number of sources which end up with negative fluxes at the final iteration and, for consistency, the \lambdar~pipeline sets these to zero also.
For the purposes of this work, we note that the fluxes for  11,210 (70.3\,per\,cent) sources are not above the $3\,\sigma$ level at 250\,$\mu$m; however, even galaxies which fall below $3\,\sigma$ do have a valid measurement and error estimate in five \herschel~bands and thus provide information for deriving dust masses. We discuss potential biases and tests in later Sections. For further details on the \lambdar~software and data release see \cite{Wright2016}.

\section{Deriving Galaxy Properties with \textsc{magphys}}
\label{sec:magphys}

For each galaxy we take the dust and stellar properties from \citet{Driver2017}, who used the \magphys\footnote{\magphys~is available from http://www.iap.fr/magphys/} package \citep{daCunha2008}  to fit model spectral energy distributions (SEDs) to the 21-band \lambdar~photometry. \magphys~uses libraries containing 50\,000 of model SEDs covering both the UV-NIR \citep{Bruzual2003} and MIR/FIR \citep{Charlot2000} components of a galaxy's SED along with a $\chi$-squared minimisation technique to determine physical properties of a galaxy, including stellar mass, dust mass and dust temperature.
\magphys~imposes energy balance between these components, so that the power absorbed from the UV-NIR matches the power re-radiated in the MIR/FIR.  In the FIR-submm regime, two major dust components are included in the libraries: a warm component (30 to 60$\thinspace$K) associated with stellar birth clouds; and a cold dust component (15$\thinspace$K to 25$\thinspace$K) associated with the diffuse ISM. A dust mass absorption coefficient of $\kappa_{850} = 0.077\thinspace\rm m^{2}kg^{-1}$ is assumed, with an emissivity index of $\beta = 1.5$ for the warm dust, and $\beta = 2$ for cold dust, where $\kappa_{\lambda} \propto \lambda^{-\beta}$.  This is consistent with the $\kappa$ values derived from observations of nearby galaxies (\citealp{James2002,Clark2016}, see also \citealp{Dunne2000}) and $\sim$2.4 times higher than the oft-used \citet{Draine2003} theoretical values (based on their $\kappa$ scaled to 850\,$\mu$m with $\beta = 2$)\footnote{We note that we have not considered the effects of changes in the dust mass absorption coefficient $\kappa$ in the different galaxy samples. As we are not able to test this using this dataset, we keep $\kappa$ constant in this work.  Different grain properties could plausibly lead to an uncertainty of a factor of a few in $\kappa$ (and therefore dust mass which scales with $\kappa$, see for example the discussion in \citealp{Rowlands2014}).}. Using the latest values for $\kappa$ in the diffuse ISM of the Milky Way from \cite{Planck2016} would give dust masses 1.6 times higher than quoted here.
For each galaxy \magphys\ uses all of the \lambdar\ measurements to find the best-fitting combination of optical and FIR model SEDs, and outputs the physical parameters for this combined SED. { We do not apply any signal to noise cuts, but low signal to noise measurements clearly do not contribute strong constraints in the fitting. So long as the estimated fluxes and uncertainties are unbiased, this makes maximum use of the information available. } 
\magphys~ also generates a `probability distribution function' (PDF) for each parameter by summing $e^{-\chi^2/2}$ over all models. The PDF for each parameter is used to determine the acceptable range of the physical quantity, expressed as percentiles of the probability distribution of model values.  The results from \magphys~for the GAMA equatorial regions are presented in \cite{Driver2016}, and we use them throughout this work. For our analysis we use the median value for each parameter, because this is more robust than the estimate from the best fit model combination. Where uncertainties are required, we use the $16^{\rm th}$ and $84^{\rm th}$ percentiles, which correspond to a $1\thinspace\sigma$ uncertainty for a Gaussian error distribution.

\begin{figure}
  \includegraphics[trim=0mm 6mm 10mm 5mm clip=true,width=\columnwidth]{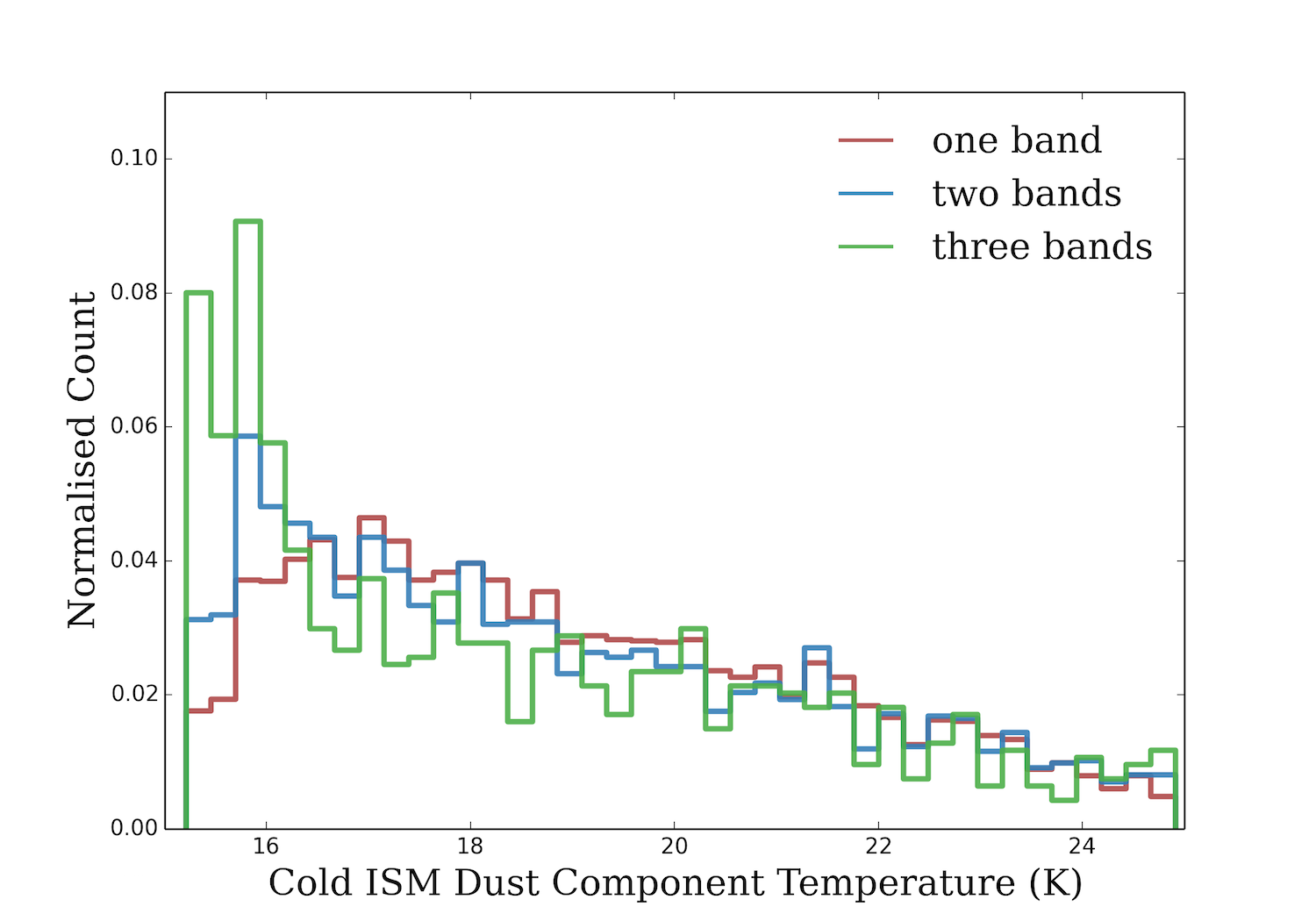}
 \includegraphics[trim=0mm 12mm 10mm -1mm clip=true,width=\columnwidth]{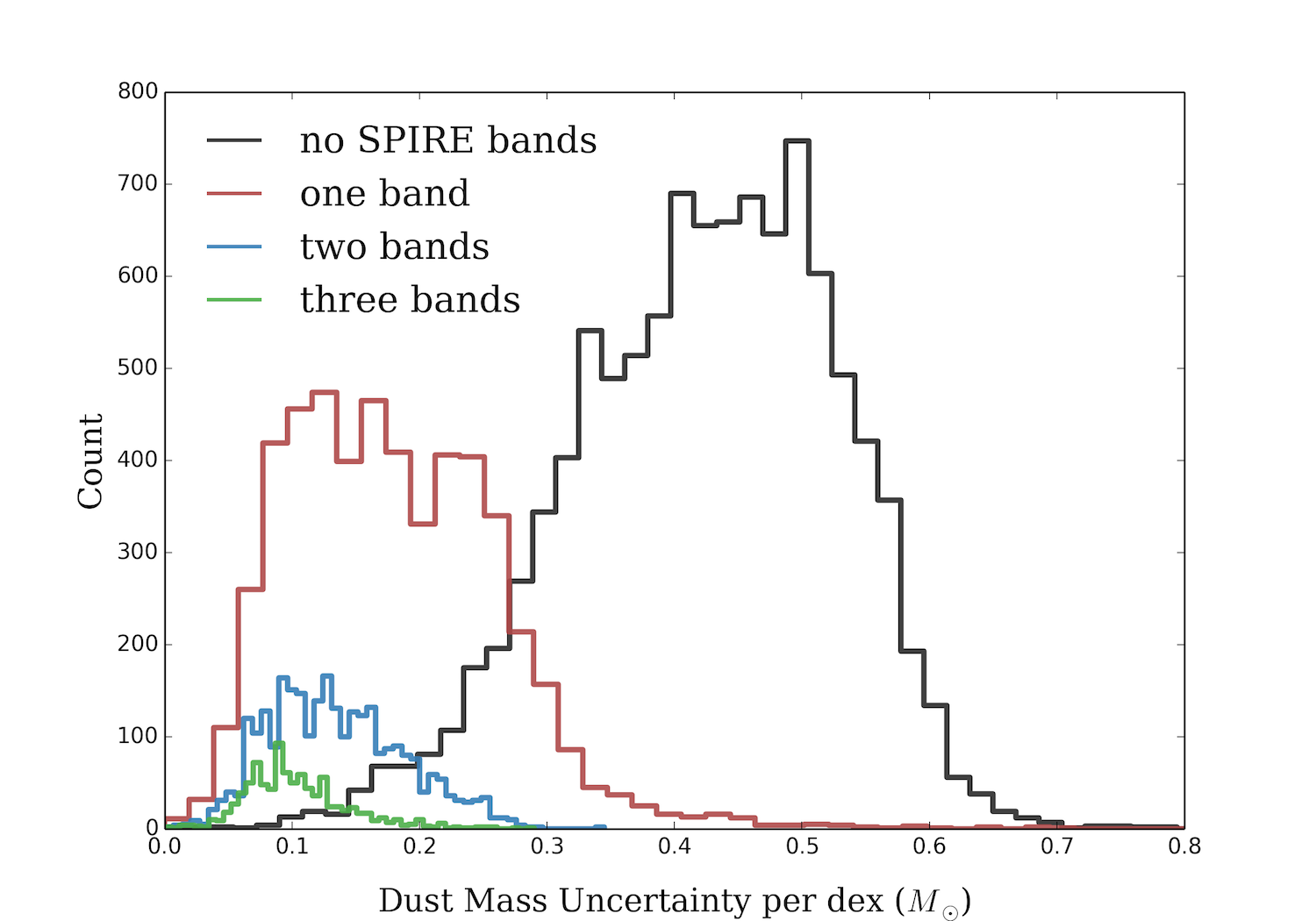}
 \caption{{\it Top:} The normalized distribution of the cold ISM dust temperature for the low redshift sample ($z\leq0.1$). The red, blue and green histograms show galaxies with $>3\,\sigma$ fluxes in one, two or three SPIRE bands respectively.  Each histogram is normalized to a total count of one: the fraction of sources in each histogram is 32, 17 and 6\,per\,cent respectively. {\it Bottom:} The distribution of uncertainties on the dust mass estimates. The uncertainties are calculated as half the difference between the 84th and 16th percentiles of the PDF; if the uncertainties are Gaussian, they correspond to one sigma. The black, red, blue and green histograms show galaxies with $>3\,\sigma$ flux measurements in zero, one, two or three SPIRE bands respectively. }
 \label{fig:dusttemphist2}
\end{figure}

\begin{figure}
 \includegraphics[trim=10mm 8mm 10mm 5mm clip=true,width=\columnwidth]{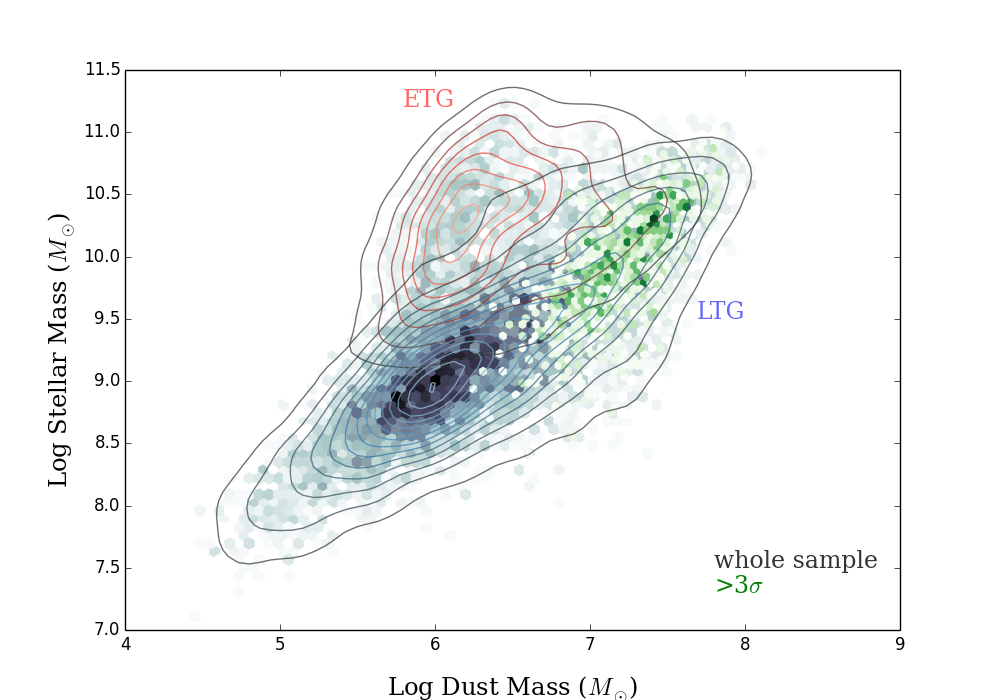}
 \caption{The distribution of dust mass and stellar mass in GAMA galaxies. The black underlying points show the whole low redshift ($z\leq 0.1$) sample. The green points show galaxies with $>3\,\sigma$ fluxes in one or more SPIRE bands. Contours show the demarcation into ETGs (black/red contours) and LTGs (black/blue contours) - see text for details. }
 \label{fig:dustmass_starmass}
\end{figure}

Our version of \magphys~is slightly modified compared to the default distribution available online. We use the most up-to-date estimates of the \herschel~band-pass profiles for both the PACS and SPIRE instruments. Also in our version, the model photometry for each of the \herschel~pass bands is calibrated to the nominal central wavelength of each band, as described in the SPIRE Handbook\footnote{The SPIRE Observer's Manual is available at http://herschel.esac.esa.int/Docs/SPIRE/spire\_handbook.pdf} \citep{Griffin2010,Griffin2013}, rather than the effective wavelength,  which is the case for other photometry. Running the code with and without these changes does not highlight any systematic error in the FIR-based \textsc{magphys} output; however, it does change individual measurements by up to a few percent.

A large fraction of the GAMA sources have measurements with signal-to-noise ratio below $3\,\sigma$ in the FIR bands: for the $z<0.1$ sample that we use here 32\,per\,cent have fluxes $>3\,\sigma$. Given that \lambdar~assigns a zero flux for each blend component that returns a negative flux at any iteration, the error distribution of faint sources becomes one-sided. { If we assume that the errors are Gaussian and consider sources which have a true flux much less than $\sigma$, then the bias introduced is the mean value of the positive half of a Gaussian i.e. $\sigma/\sqrt{2\pi} \approx 0.4\,\sigma$. Sources with more positive fluxes will have a smaller bias.}

\subsection{Temperatures} 

The normalized distribution of dust temperatures output by \magphys~for {\sc lambdar} sources with fluxes above $3\,\sigma$ in one, two or three {\it Herschel}-SPIRE bands is shown in Fig.~\ref{fig:dusttemphist2} (top panel).  Where we have sources with \herschel~fluxes $>3\,\sigma$ in one or more bands, the temperature is well constrained ($\pm \sim 1\,{\rm K}$), and has a tendency to be fairly cold, $\sim 18$\,K. There is also a tendency for the galaxies with \herschel~fluxes $>3\,\sigma$ in all three bands to be colder than those with only one or two bands; this is not unexpected given that the combination of the shape of the SED of a modified blackbody, and the more sensitive bluer SPIRE bands.  The temperature histogram for these sources appears to continue to rise at temperatures below $17$\,K, with a peak at $16$\,K.  This potentially suggests that a colder dust prior than the $15-25$\,K used in this work might be needed for a small fraction of galaxies (e.g. \citealt{DeVis2017a,Viaene2014,Smith2012}). We will return to this below.

For the galaxies that have fluxes below $3\,\sigma$ in all of the \herschel~SPIRE bands we have poor constraints on the cold dust temperature. For these galaxies, the temperature PDF follows the underlying \emph{flat} temperature prior used in the \magphys~code with limits from 15-25\,K. { Since the temperature estimate is the median of the PDF, this tends towards the median of the prior as the constraints become weaker\footnote{In Appendix \ref{Appendix:Temperature}, we test if this leads to any bias in our DMF, and conclude that there is no significant bias.}.}  Despite this, the combination of UV and optical photometry and \emph{the FIR measurements} do provide useful information on the dust masses for those galaxies with FIR fluxes $<3\,\sigma$ in all Herschel bands. This can be seen in Fig.~\ref{fig:dusttemphist2} (bottom panel), which shows the distribution of estimated dust mass uncertainties for galaxies with $>3\,\sigma$ in zero, one, two or three SPIRE bands. For the subsets in one, two or three bands the corresponding uncertainties in mass are 0.18, 0.14 and 0.1\,dex. Galaxies with $<3\,\sigma$ in any SPIRE band typically have dust mass uncertainties of 0.4\,dex on average.

As a further, though indirect, check that the estimated uncertainties are reasonable we look at the distribution of dust mass and stellar mass of the GAMA $z\le 0.1$ sample, as shown in Fig.~\ref{fig:dustmass_starmass}. The sources with fluxes $>3\sigma$ in at least one band are shown in green (as expected, these are the more dusty galaxies), with the entire sample shown by the grey bins. We see that the distribution shows a marked bimodality in this plane, clearly visible even for sources without fluxes $>3\sigma$ in any of the FIR bands.  To investigate this further, Fig.~\ref{fig:dustmass_starmass} highlights the morphological classifications of the galaxies, split into ETGs and LTGs \citep{Moffett2016_gals}\footnote{Here we have not included the little blue star-forming spheroids.}.  The ETGs have many fewer $>3\,\sigma$ sources than the LTGs, even for bright optical sources, and this is as expected given that ETGs contain an order of magnitude less dust than late-type galaxies of the same stellar mass (see e.g. \citealp{Bregman1998,Clemens2010,Skibba2011,Rowlands2012,MWLSmith2012,Agius2013,Agius2015}). If the true uncertainties in $M_{\rm d}$ were larger than 0.5\,dex, the bimodal structure in Fig.~\ref{fig:dustmass_starmass} would be smeared out, suggesting the errors in \magphys\ do reasonably represent the uncertainties.

\subsection{ Dust masses and the temperature prior} 

The cold dust temperature prior is clearly going to impose some limits on the dust mass uncertainty from the fits. However, we argue that the prior temperature range from \magphys~used in this work is appropriate for a number of reasons. (i) A range of cold dust temperatures between 15-25\,K is in fact a good description of the observed range of cold dust temperatures in galaxies \citep{Dunne2001,Skibba2011,SmithMLW2012,Clemens2013,Clark2015}. (ii) \citet[Appendix~A]{Smith2012} investigated whether a broader temperature prior should be used in \magphys~fitting. They found that changing the prior range suggested that only 6\,per\,cent of their \herschel~detected sources were actually colder than 15\,K.  They also demonstrated that adopting a wider temperature prior is not always appropriate given the non-linear increase in dust mass when the temperature falls below 15\,K (where the SPIRE bands are no longer all on the Rayleigh-Jeans tail). At T<15\,K, symmetrical errors in the fitted temperature produce a very skewed PDF for the dust mass and result in a population bias to higher dust masses for a distribution of Gaussian errors in cold temperature. Furthermore, in relation to SED fitting, a very cold dust component contributes very little to the luminosity in the FIR per unit mass, so it can be included by a fitting routine with very little penalty in $\chi^2$ when the photometry in the FIR and sub-mm is of low SNR. Indeed \citet{Smith2012} use simulated photometry to show that galaxy dust masses can be overestimated by (in excess of) 0.5\,dex when widening the prior to below 15\,K; they therefore strongly caution in using wider temperature priors for sources with weak sub-mm constraints (as is the case here).  (iii) Though some galaxies have been shown to require colder dust temperatures than 15\,K (\citealp{Viaene2014,Clark2015,DeVis2017a}, Dunne et al. {\it submitted}), { the fraction of our sample with $>3\,\sigma$ in at least one band that have dust temperatures $<16\,\rm K$ is $<9$\,per\,cent.}

As an example to illustrate the potential size of the effect, consider the case that 6\,per\,cent of our galaxies had a true dust temperature of 12\,K but instead we fit a temperature of 15\,K due to the limited prior. We would underestimate the dust mass for this population by a factor of $\sim 2.6$ (ie 0.4\,dex). However, 94\% of galaxies have true temperatures in the range 15--25\,K and since most of them do not have $>3\sigma$ FIR fluxes they will have errors on the fitted temperature of order $\pm5$\,K. Widening the prior to extend to 12\,K would mean that 16\,per\,cent of sources would be erroneously returned a temperature which was below 15\,K resulting in a large positive bias to their dust masses. 

{  Appendix \ref{Appendix:Temperature} presents a more
  thorough investigation of the effects on the DMF that result from
  poorly constrained cold dust temperatures for galaxies with low
  signal to noise in the FIR. }

\section{The Dust Mass Function}
\label{sec:dmf}

\subsection{Volume Estimators}
\label{sec:dmf_volEst}
{ To estimate the dust mass function, we use the $V_{\rm max}$ method
\citep{Schmidt1968} with a correction to account for density
fluctuations as suggested by \cite{Cole2011}.}

\begin{equation}
\phi(M_i) = \sum\limits_{n=1}^{N_i} \left(\frac{1}{V'_{\rm
      max,n}}\right) =
\sum\limits_{n=1}^{N_i} \left(\frac{1}{V_{{\rm eff},n}} \frac{\langle
\delta_f \rangle}{\delta_n}\right) ,
\label{eq:dmf_sum}
\end{equation}

\noindent where $V_{{\rm eff},n}$ is the effective volume accessible to a galaxy { within the redshift range chosen}, and the sum extends over all $N_i$ galaxies in the bin $M_i$ of the mass function; $V'_{\rm max}$ is the density-corrected accessible volume; $\delta_n$ is the local density near galaxy $n$, as defined below; and $\langle \delta_f \rangle$ is a fiducial density for each field, also defined below.

We use two methods to estimate the accessible volume for each galaxy. First we derive $V_{\rm max}$ for each galaxy by estimating the maximum redshift at which that source would still be visible given the limiting magnitude of the survey. This requires taking into account both the optical brightness of each galaxy and the $K$-correction required as the galaxy SED is redshifted. { The maximum redshift is not allowed to exceed the user-imposed redshift range of the sample (here we use $z<0.1$).} Using this maximum redshift and the area of the survey, an accessible comoving volume can be calculated. These maximum volumes are the same as used in W17. We refer to this method as $pV_{\rm max}$, since it is based on the simple  photometric selection of the survey.

The second method we use to estimate the $V_{\rm max}$ for each galaxy is based on a bivariate brightness distribution (BBD). This involves binning the data in terms of the two most prominent selection criteria, and aims to account for the selection effects that they introduce. Since our sample is optically selected, we choose the absolute $r$-band magnitude, and for the second axis we choose surface brightness in the $r$-band { \citep{Loveday2012,Loveday2015}}. We have estimated fluxes in all other bands for all galaxies, even if they are not significantly detected, so we do not directly apply any further selection criteria.

\begin{figure*}
\begin{center}
 \includegraphics[trim=0mm 4mm 9mm 5mm clip=true,width=0.8\textwidth]{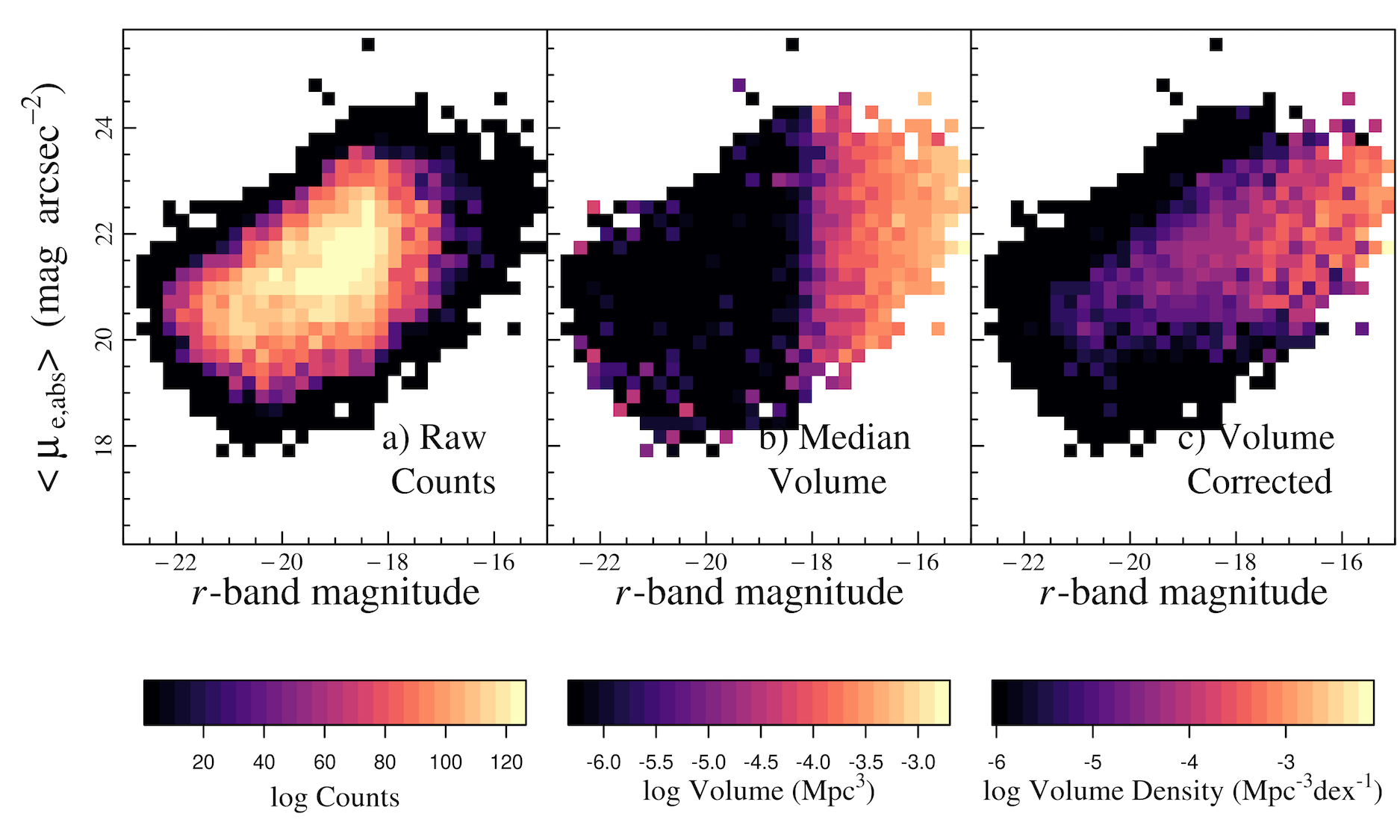}
 \caption{The bivariate brightness distribution (BBD) for our sample with surface brightness and $r$-band magnitude as the two ``axes" (W17) with a) Raw counts in surface brightness/$r$-band magnitude bins, b) Median volume in surface brightness/$r$-band magnitude bins, c) Weighted counts, i.e. volume density in the surface brightness/$r$-band bins. { Each of the panels represents the BBD resulting from the median of 1000 Monte Carlo simulations where we perturb the $r$-band magnitude and surface brightness within their associated uncertainties}.}
 \label{fig:DM_bbd}
\end{center}
\end{figure*}

\begin{figure}
 \includegraphics[trim=10mm 5mm 0mm 0mm clip=true,width=\columnwidth]{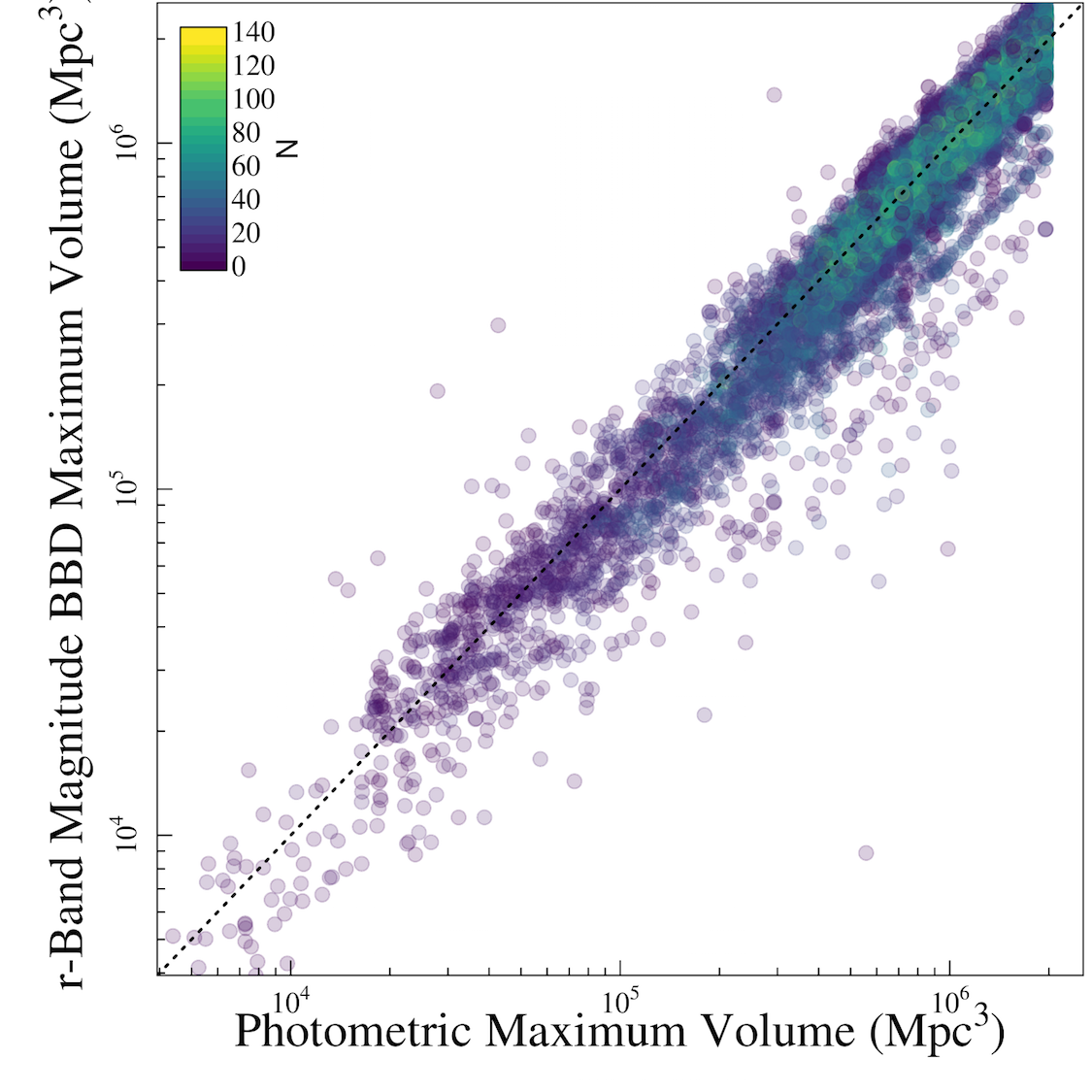}
 \caption{The maximum effective volumes for our galaxies at $z<0.1$ derived using the $pV_{\rm max}$ method (x-axis), and BBD method using $r$-band magnitude and surface brightness as the two selection features (y-axis). The colour of the points is determined by the number of galaxies in the BBD bin that each galaxy resides in (Fig \ref{fig:DM_bbd}), as shown by the colour bar in the top left corner. { We note that the number of galaxies per bin is the median resulting from 1000 Monte Carlo simulations, where we perturb the $r$-band magnitude and surface brightness within their associated uncertainties.}}
 \label{fig:BBDVmaxComp}
\end{figure}

This method follows closely the format of the Galaxy Stellar Mass Function (GSMF) produced by W17 for the same sample; see also Fig. \ref{fig:DM_bbd}, which is a diagrammatic representation of the BBD method.  For each 2D $r$-band magnitude/surface brightness bin (Fig. \ref{fig:DM_bbd}a), the volume enclosed by the median luminosity distance of the galaxies in the bin and the on-sky area of GAMA is calculated (Fig. \ref{fig:DM_bbd}b) and doubled in order to find an `accessible volume' for all of the galaxies in that bin (Fig. \ref{fig:DM_bbd}c).  Using twice the median value will provide an effective $V_{\rm  max}$ that, at some level, corrects for the incompleteness at large distances whatever the cause of the incompleteness.  Thus the BBD method has the benefit that it can correct for selection effects in two parameters at once. Using the median volume to determine the effective $V_{\rm max}$ has the advantage that it is more statistically robust than the actual maximum volume observed in a given bin. However, this estimator is only strictly valid when the underlying galaxy distribution in any given bin is randomly and evenly distributed in space, so the average $V/V_{\rm max} = 0.5$.  Given the large density fluctuations seen in the galaxy distribution, we cannot state that it is always the case, particularly for local, low-mass galaxies, which are hampered by small-number statistics and strongly affected by cosmic variance. It is more likely to be the case that $V/V'_{\rm max} = 0.5$, i.e. the maximum volume weighted by density. To allow for these density fluctuations, we find a median weighted by the inverse of the density correction factors $\delta_n/\langle \delta_f\rangle$, defined below. Galaxies in over-dense regions are given less weight in the median compared to galaxies in under-dense regions, so any bias in the median volume from density fluctuations should be minimised. { We note that in order to reduce noise introduced into the DMF from BBD bins with poor statistics we perform a Monte Carlo (MC) simulation whereby we perturb the quantities used for the two `axes' of our BBD within their associated uncertainties and recalculate the BBD 1000 times and find the median BBD $V_{\rm max}$ associated with each bin. In essence, this smooths the BBD by the estimated errors, and reduces the uncertainty in the BBD $V_{\rm  max}$.}

A direct comparison of the maximum volumes derived from both the $pV_{\rm max}$ and BBD methods is shown in Fig. \ref{fig:BBDVmaxComp} with the points coloured by the average number of galaxies in the BBD bin containing that galaxy across all the MC simulations. The largest deviation from the 1:1 line is seen for galaxies that lie in bins with a small number of galaxies contributing to the median volume. These volumes are generally low, meaning they are also strongly affected by cosmic variance. The $pV_{\rm max}$ values are systematically higher by 0.8\% on average than those derived from the BBD method, { which translates to an average offset of 1\% in the binned DMF values when determined by the median weighted by the error on the measurement.}


Since we compare to the galaxy stellar mass function from W17, who use stellar mass and surface brightness as the BBD axes, it may be argued we should use the same approach. We consider this  in Appendix \ref{Appendix:axes}, and conclude that the Schechter parameters are consistent with the $r$-band and surface brightness BBDs within uncertainties. We opt to use the $r$-band magnitude for our second axis here as it is more in line with the optical $pV_{\rm max}$, and does not depend on stellar properties directly.

\begin{figure}
 \includegraphics[width=\columnwidth]{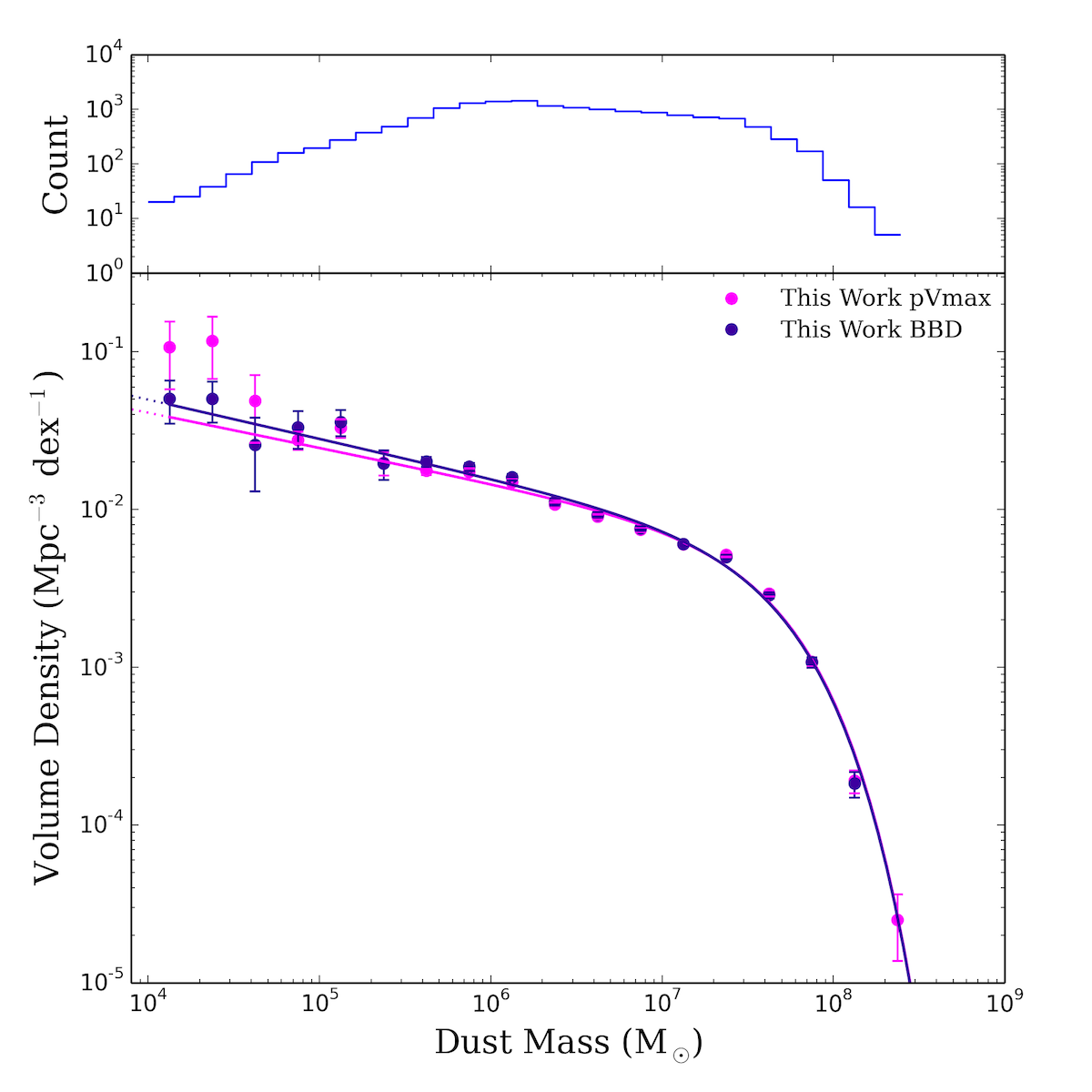}
 \caption{The $pV_{\rm max}$ (purple) and BBD (blue) dust mass functions for the GAMA/{\it H-}ATLAS sources at $z<0.1$.  The data points show the observed values corrected for over and under densities in the GAMA fields (see W17). The solid lines are the best fitting (minimum $\chi^2$) single Schechter functions from our SB measurements. Error bars are derived from our PB measurements. The total number of sources in each bin is shown in the top panel.}
 \label{fig:dmf}
\end{figure}

Density fluctuations in the GAMA equatorial fields e.g. \citet{Driver2011,Dunne2011} have a pronounced effect on the DMF, and so we apply density corrections as calculated by W17 to account for the over- or under-densities present in each of the equatorial fields (see e.g. \citealt{Loveday2015}). These multiplicative corrections were derived as a function of redshift by determining the local density of the survey at the redshift of the galaxy in question.  This is achieved by simply finding the running density as a function of redshift, and convolving this function with a kernel of width 60\,Mpc.  These were compared to the fiducial density, taken from a portion of the GAMA equatorial fields with stellar masses above $10^{10}\, M_{\odot}$ and $0.07 < z<0.19$. This subset was chosen because of its high completeness level, uniform density distribution, and low uncertainty due to cosmic variance.  To correct the effective volume for galaxy $n$, $V_{{\rm eff}, n}$, we simply multiply by a factor of $\delta_n / \langle \delta_f \rangle$ to obtain $V'_{\rm max}$.

To remove any spuriously low $V'_{\rm max}$ values introduced either by the density correction factor or by uncertainties in the calculated $V'_{\rm max}$, we employ a clipping technique. We split the galaxies into 100 stellar mass bins and remove 5\% of the most spurious $V'_{\rm max}$ values, and up-weight the remaining galaxies accordingly  { giving a final sample size of 15,750. For consistency with W17, the 5\% clipping is performed on the total sample, i.e. before the imposition of the requirement of {\it H-}ATLAS coverage, translating to the removal of $\sim200$ galaxies from the sample requiring \textit{H-}ATLAS coverage that we use for this work. W17 perform a one-sided clipping, since higher $V'_{\rm max}$ values tend to be more stable than lower ones since brighter galaxies tend to have better constraints. Galaxies with high $V'_{\rm max}$ values also contribute less volume density and therefore tend to add less noise to their given bin than faint galaxies. } Once removed, the weights of the remaining galaxies are scaled by the fraction of removed galaxies\footnote{This has the effect of smoothing the low-mass end of the DMF.}.

\subsection{The Shape of the DMF}
\label{DMFshape}

The DMF, derived for the largest sample of galaxies to date, based on the optically selected GAMA sample, is shown in Fig.~\ref{fig:dmf} using the two methods described in Section~\ref{sec:dmf_volEst} to calculate volume densities.  We have extended the function well below the low dust mass limit of all previous studies; indeed we extend to dust masses $\sim 10^{4}\, M_{\odot}$ whilst dust masses above $10^{4.5}\, M_{\odot}$ are well constrained.  We have therefore extended the observed range of the DMF by $\sim$2\,dex in $\rm M_{d}$ { compared to e.g. \citet{Dunne2011} and significantly reduced the statistical uncertainty compared to previous measurements (with $\sim$70$\times$ the sample size, see Section~\ref{sec:otherdmfs} for more details).} The offset at the low-mass end of the DMF seen between the two methods can be attributed to the differences shown in Fig. \ref{fig:BBDVmaxComp}, the sources with the lowest dust mass tend to be those which are nearby and faint, and so most likely to be affected by small number statistics when calculating the BBD $V_{\rm max}$ values.

\begin{tiny}

\begin{table*}
\centering
    \begin{tabular}{|l|c|c|c|c|c|}
    \hline \hline
    Survey   			&		 		$M^{*}$  	& 		 $\alpha$  			& $\phi^{*}$  & $f_{\rm mix}$ & $\Omega_{\rm d}$  \\
		& ($10^7\,h^2_{70}\,\rm M_{\odot}$) &    & ($10^{-3}\,h^3_{70}\,\rm Mpc^{-3}\,\rm dex^{-1}$) &  & ($10^{-6}$)\\ \hline
			C13         &  $\ClemMstar$ & $\Clemalpha$ & $\Clemphistar$ & - &$\ClemOmegad$ \\ \\
			D11          & $\DunneMstar$ & $\Dunnealpha$   & $\Dunnephistar$ & - & $\DunneOmegad$  \\ \\
			V05      & $\VlakMstar$ & $\Vlakalpha$ &  $\Vlakphistar$   & - & $\VlakOmegad$ \\ \\ \hline
    This work single $pV_{\rm max}$  & $\pVmaxMstar$  & $\pVmaxalpha$    &  $\pVmaxphistar$ & - & $\pVmaxOmegad$\\
    This work single BBD     &      $\bbdDMMstar$ 	 &   $\bbdDMalpha$  	    & $\bbdDMphistar$ & - & $\bbdDMOmegad$ \\ \\
    Deconvolved single $pV_{\rm max}$  & $\pVmaxConMstar$  & $\pVmaxConalpha$    &  $\pVmaxConphistar$ & - & $\pVmaxConOmegad$\\
    Deconvolved single BBD     &      $\bbdDMConMstar$    &   $\bbdDMConalpha$        & $\bbdDMConphistar$ & - & $\bbdDMConOmegad$\\ \hline \hline
     &  (1) , (2) &  (1) , (2) & ,  & \\ \hline
    { This work DSF $pV_{\rm max}$}  & ($\pVmaxDSFMstarone$) , ($\pVmaxDSFMstartwo$) & ($\pVmaxDSFalphaone$) , ($\pVmaxDSFalphatwo$)  & $\pVmaxDSFphistar$ & $\pVmaxDSFfmix$ & $\pVmaxDSFOmegad$ \\
    { This work DSF BBD}     &  ($\bbdDMDSFMstarone$) , ($\bbdDMDSFMstartwo$) & ($\bbdDMDSFalphaone$) , ($\bbdDMDSFalphatwo$)  & $\bbdDMDSFphistar$ & $\bbdDMDSFfmix$ & $\bbdDMDSFOmegad$  \\ \\
    Deconvolved  DSF $pV_{\rm max}$  & ($\pVmaxConDSFMstarone$) , ($\pVmaxConDSFMstartwo$) & ($\pVmaxConDSFalphaone$) , ($\pVmaxConDSFalphatwo$)  & $\pVmaxConDSFphistar$ & $\pVmaxConDSFfmix$ & $\pVmaxConDSFOmegad$ \\
    Deconvolved  DSF BBD    &   ($\bbdDMConDSFMstarone$) , ($\bbdDMConDSFMstartwo$) & ($\bbdDMConDSFalphaone$) , ($\bbdDMConDSFalphatwo$)  & $\bbdDMConDSFphistar$ & $\bbdDMConDSFfmix$ & $\bbdDMDSFOmegad$  \\ \hline \hline
    \end{tabular}
        \caption{Schechter function values for dust mass functions in the literature and this work for both the SSF and DSF fits).  The other literature studies include: C13 - \citealp{Clemens2013}, D11 - \citealp{Dunne2011}, V05 - \citealp{Vlahakis2005}. All have been scaled to the same dust mass absorption coefficient used here.  The \citet{Dunne2011} DMF includes a correction of 1.42 for the density of the GAMA09 field \citep{Driver2011} and the fits in this work include the density-weighted corrections from W17.  For comparison we include the deconvolved Schechter function fit parameters in the final section of the table (see Section~\ref{sec:deconvolve}), which are very similar to the ordinary Schechter function parameters. { We also include the double Schechter function (DSF), and deconvolved DSF with their major component and minor component listed under (1) and (2) respectively for each non-coupled SF parameter (see Equation \ref{eq:double}). } }
    \label{tab:schechterTab}
\end{table*}

\end{tiny}

We estimate uncertainties on the volume densities calculated here using three techniques. { First, using a jackknife method in two different ways: (i) taking random subsamples of the data, and (ii) by splitting the sample by on-sky location. Second, we perform 1000 bootstrap resamplings on our volume densities to determine the sample errors. We refer to this as the simple bootstrap or SB method.  Third, we use the bootstrap technique but for each realisation, we also perturb each dust mass by a Gaussian random deviate with $\sigma$ set according to the 16-84 percentile uncertainty from {\sc magphys} (hereafter the PB method).}  Unsurprisingly, Poisson noise estimates agree with all these techniques at the high mass end ($M_{\rm d}>10^{7.5}\, M_{\odot}$), but underestimate the uncertainty in the low dust mass bins ($M_{\rm d}<10^6\, M_{\odot}$). The random jackknife and SB error estimates agree very well (within 0.5\,\%), whereas the on-sky jackknife uncertainty is around 5\,\% higher. This is not unexpected since this method will include a component of uncertainty from cosmic variance within the survey volume. By disentangling the statistical uncertainty from the cosmic variance uncertainty, the larger uncertainty in the on-sky jackknife suggests an error due to cosmic variance of at least 7\,per\,cent assuming that the difference is due only to cosmic variance. The cosmic variance estimator from \cite{Driver2010}\footnote{\url{cosmocalc.icrar.org}} suggests an error of 16.5\,per\,cent for the full survey volume. This is significantly higher than the effective cosmic variance that we measure, because we make corrections for the density variations within the survey volume.  { For the rest of this work, we use the simple bootstrap method without perturbation of the dust mass (SB) for the data points. For the uncertaintoes we use the bootstrap with additional perturbation using the {\sc magphys} dust mass uncertainties (PB) since this takes into account both the variation within the sample and the uncertainty in the dust mass estimations themselves. {  As discussed in section \ref{sec:deconvolve} the PB is likely to give biased estimates of the best fit parameters, but since it includes our mass uncertainties, it provides a better estimate of the uncertainties on the best fit parameters. } 
}

Following \citet{Dunne2011}, we { fit a single Schechter function (SSF)} \citep{Schechter1976} to the observed DMF, using $\chi^2$ minimisation to derive the best-fit values for $\alpha$, $M^{*}$ and $\phi^{*}$ which are the power law index of the low-mass slope, the characteristic mass (location of the function's `knee'), and the number volume density at the characteristic mass respectively.  This takes the form (in ${\rm log}M$ space):

\begin{align}
  S(M; \alpha, M^{*}, \phi^{*}) & = \phi^{*} e^{-10^{{\rm log}M - {\rm log}M^*}} \nonumber \\
 & \times \left(10^{{\rm log}M - {\rm log} M^*}\right)^{\alpha+1} d\,{\rm log}M,
\end{align}

\noindent where we have explicitly included the factor ${\rm ln} 10$ in the definition of $\phi^{*}$, such that $\phi^{*}$ is in units of $\rm Mpc^{-3}\,dex^{-1}$.

{ We  fit a Schechter function to each of our bootstrap realisations, and use the median of the resulting values as the best fit value for each parameter. We use the standard deviation between the values to estimate uncertainty on the parameters. The parameters for both the $pV_{\rm max}$ and BBD fits are quoted in Table~\ref{tab:schechterTab}. Note that cosmic variance will introduce further uncertainty in our measurements. This will mostly be seen as an increased uncertainty on $\phi^*$, though both $M^*$ and $\alpha$ will also have slightly larger errors. }

\begin{figure*}
 \includegraphics[trim=10mm 5mm 0mm 0mm clip=true,width=\columnwidth]{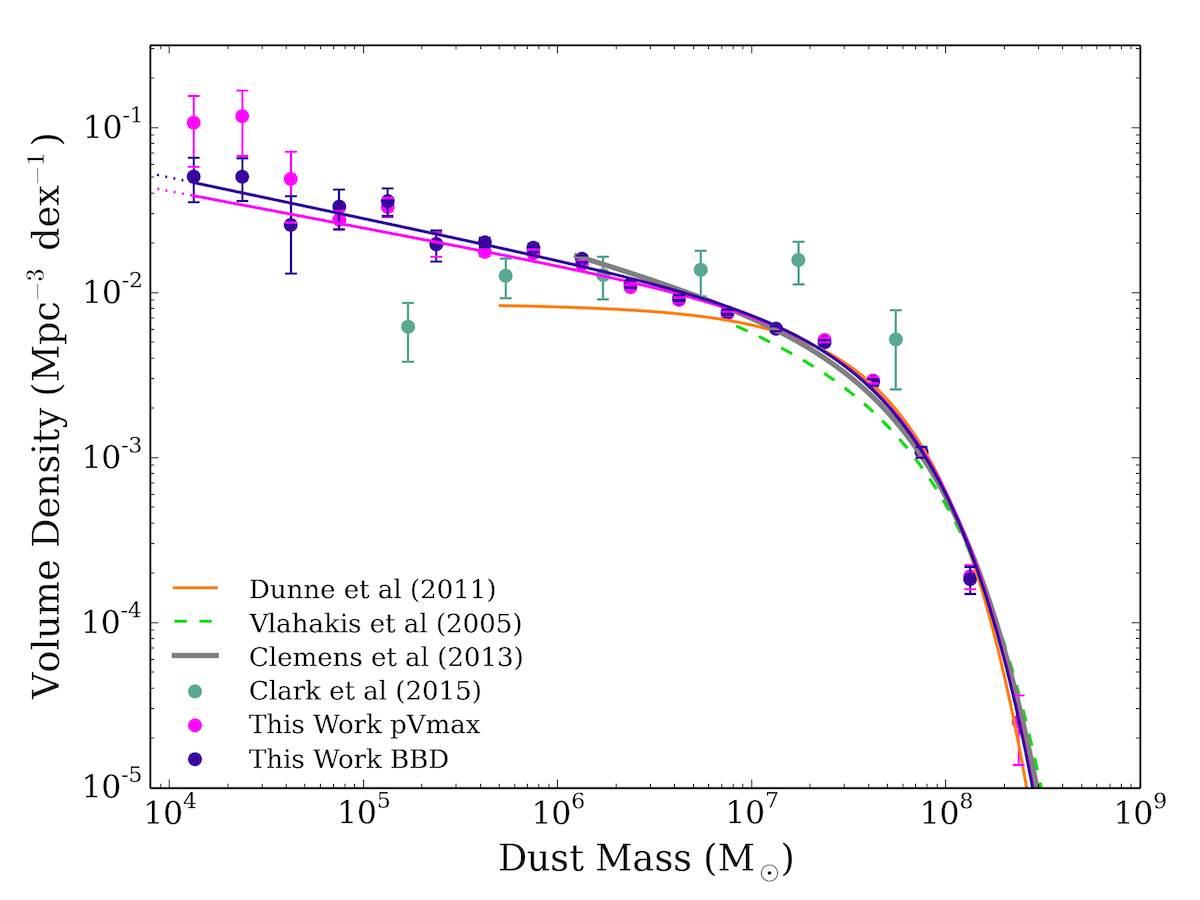}
 \includegraphics[trim=9mm 10mm 8mm 0mm clip=true,width=\columnwidth]{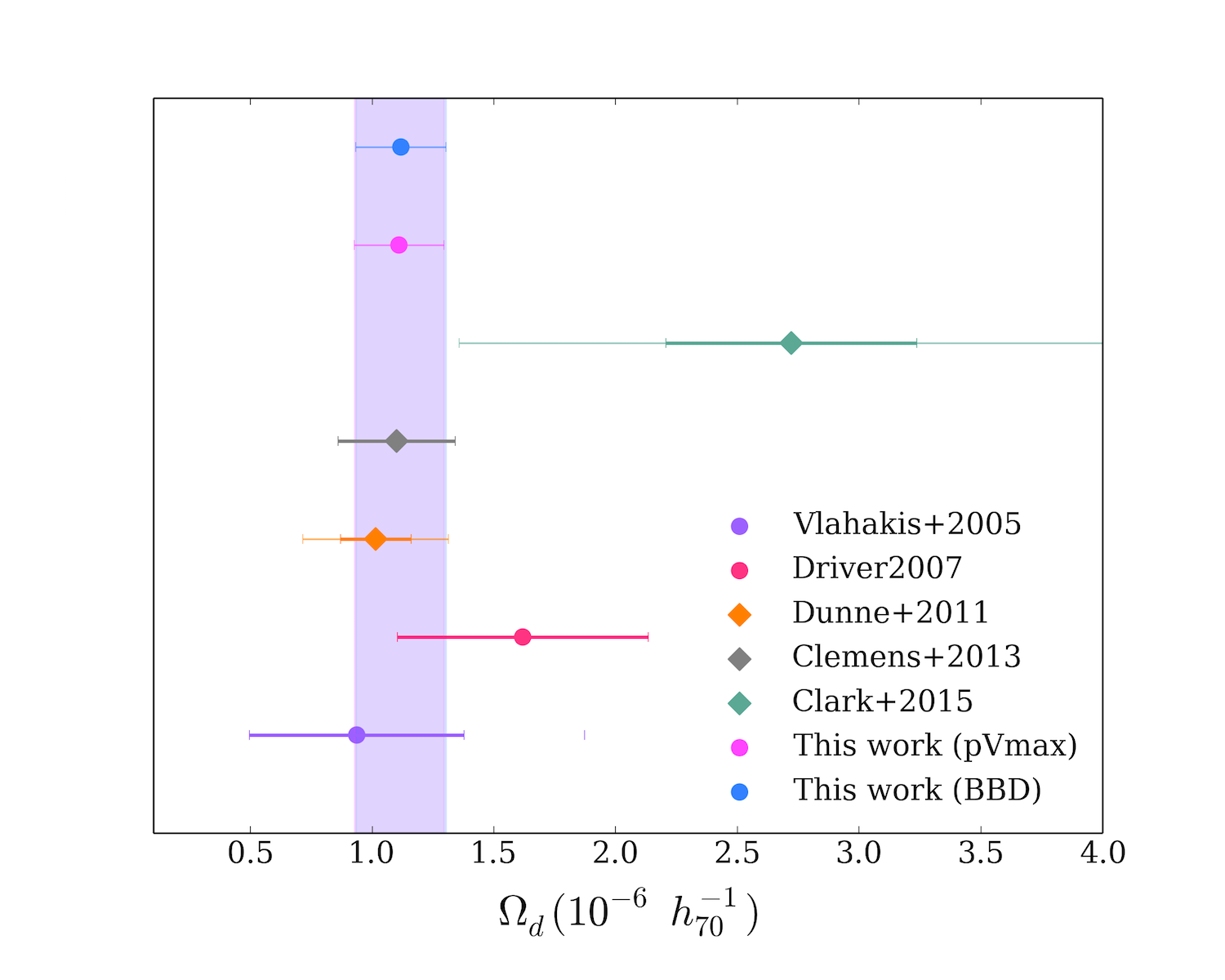}
 \caption{ Comparison of the ({\it left}) DMF and ({\it right}) dust mass densities $\Omega_{\rm d}$ from this work with those from the literature.  We compare with (i) the blind, local $z<0.01$ galaxy sample from \citet{Clark2015} (ii) the all-sky local star-forming galaxies from the bright \textit{Planck} catalogue from \citet{Clemens2013} (iii) the ground-based submm measurements of local optical galaxies from \citet{Vlahakis2005} and (iv) the 222 galaxies out to $z<0.1$ from the {\it H-}ATLAS survey \citep{Dunne2011}.  Schechter fit parameters are listed in Table~\ref{tab:schechterTab}.  The dust density parameter ($\Omega_{\rm d}$) measurements are scaled to the same cosmology, with diamonds representing dust-selected measurements and circles representing optically-selected samples. Our work are shown as $pV_{\rm max}$ and BBD for the single Schechter fit - SSF to each. The solid error bars on $\Omega_{\rm d}$ indicate the published uncertainty derived from the error in the fit whilst the transparent error bars indicate the total uncertainty derived by combining the published uncertainty and the cosmic variance uncertainty estimate for that sample (where known). We note that the solid error bars indicating the uncertainty from our bootstrap analysis lie within the point itself for both our BBD and $pV_{\rm max}$ values.   The \citet{Dunne2011} DMF includes the correction factor of 1.42 for the density of the GAMA09 field \citep{Driver2011} whilst our data points have been weighted by density correction factors from W17.  The shaded region emphasises the range of $\Omega_{\rm d}$ derived from our { observed SSF fits} to the DMFs with width showing the error from cosmic variance.  }
 \label{fig:dmf_others}
\end{figure*}

\begin{figure*}
\includegraphics[trim=10mm 10mm 10mm 5mm clip=true,width=0.95\textwidth]{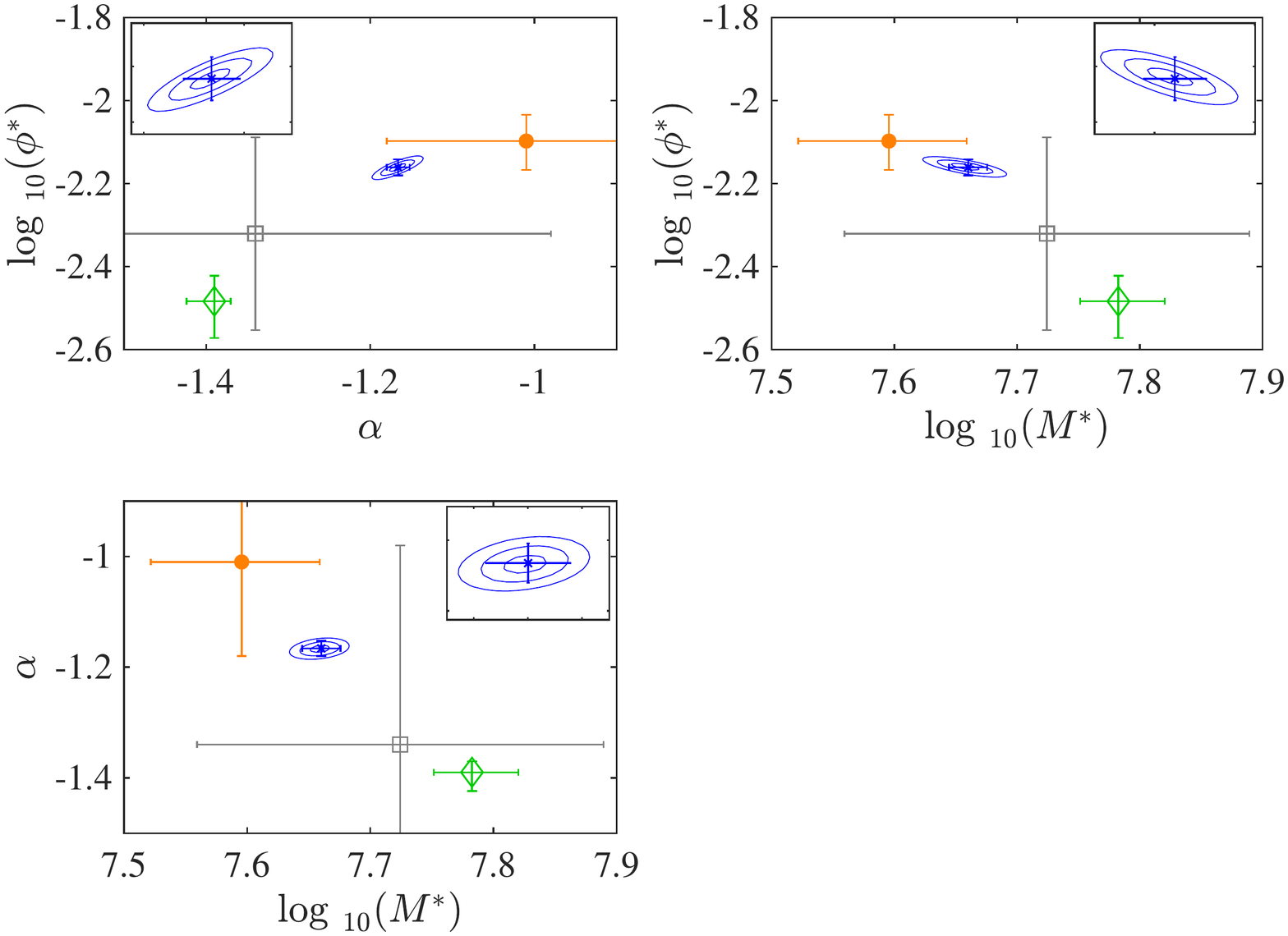}
 \caption{{ The confidence intervals for the $pV_{\rm max}$ single Schechter dust mass function fit parameters derived in this work (blue ellipses) showing the correlation between the fit parameters (insets) and comparison with previous values (note that $\phi^*$ is in units of $\rm Mpc^{-3} dex^{-1}$).  Error bars on our fit parameters are taken from the $\Delta \chi^2=1$ for each parameter (these are consistent with errors derived from the bootstrap process described in Section~\ref{sec:dmf}). The contours are from the 1, 2, 3 $\sigma$ values of $\Delta \chi^2$ for the parameter slice centred on the best fit for the non-plotted 3rd parameter.} Green denotes \citet{Vlahakis2005}, orange represents \citet{Dunne2011}, and grey shows \citet{Clemens2013}. We note that the error bars on the \citet{Vlahakis2005} values were derived using Poisson statistics, and so may be an underestimate of the error in the measurements. }
 \label{fig:conf_intervals}
\end{figure*}

\subsection{Comparing the Dust Mass Function with previous work}
\label{sec:otherdmfs}

{ We compare the SSF parameters derived here with single Schechter function fits in the literature} (Fig.~\ref{fig:dmf_others} left and Table~\ref{tab:schechterTab}). We also compare the confidence intervals for our derived parameters in Fig.~\ref{fig:conf_intervals} with previous work.  For the first time we are able to directly measure the functional form at masses below $5 \times 10^5\, M_{\odot}$ and determine the low mass slope of the DMF, $\alpha$.  We see that there is a good overall match at the high mass end with the \citet{Dunne2011} DMF, but at the faint end, the DMF is steeper than predicted from the \citet{Dunne2011} function suggesting larger numbers of cold or faint galaxies than expected.  We note that the \citet{Dunne2011} sample is different to our DMF in two ways (i) it is a dust-selected (or rather 250-$\mu$m-selected) sample rather than optically selected and (ii) was drawn from the {\it H-}ATLAS science demonstration phase data, which is only 16\,sq\,deg of the GAMA09 field at $z<0.1$ and is known to be under-dense compared to the other GAMA fields \citep{Driver2011}. Our DMF is also similar to the optically-selected \citet{Vlahakis2005}\footnote{Here we quote the PSCz-extrapolated DMF from \cite{Vlahakis2005} where they assume a 20$\,$K cold dust component for their sources.} SSF at the highest masses, though we find a higher space density of galaxies around the `knee' of the function potentially due to the higher redshift limit probed in this study and improvement in statistics in this work. In general, the 2-d parameter comparisons in Fig.~\ref{fig:conf_intervals} show that the DMF in this work has intermediate values of $\alpha$, $M^{*}$ and $\phi^{*}$ in comparison to the \cite{Clemens2013}\footnote{The fit parameters quoted in this work for \citet{Clemens2013} are different to those that appear in their paper and in \citet{Clark2015}. The reason for this is that \citet{Clemens2013} did not include the $\mathrm {ln}{10}$ factor when calculating their integrated dust densities, and in \citet{Clark2015} we erroneously attributed this error to a missing per dex factor in $\phi^{*}$. In fact their error was only in converting from $\phi^{*}$ to $\rho_{\rm d}$.}, \cite{Vlahakis2005}, and \cite{Dunne2011} parameters but here we have tighter constraints due to the larger sample of sources. Differences could also arise because of the variation of best-fit parameters with the minimum mass limit of the fit since all the surveys have different mass ranges. We discuss the implications of changing the minimum mass limit of our fits in Appendix \ref{Appendix:mmin}.

The integrated dust mass density parameter $\Omega_{\rm d}$ at $z \le 0.1$ is derived by using the incomplete gamma function to integrate down to $M_{\rm d} = 10^{4}\,M_{\odot}$ (our lower limit on measurement of the form of the DMF). This gives $(\pVmaxOmegad) \times 10^{-6}$ for both the $pV_{\rm max}$ and BBD methods. For comparison, our $\Omega_{\rm d}$ values calculated without imposing this limit are $(\pVmaxOmegadinf) \times 10^{-6}$ and $(\bbdDMOmegadinf) \times 10^{-6}$ for the $pV_{\rm max}$ and BBD methods respectively, so the difference is very small. Previous measurements of $\Omega_{\rm d}$ are shown in Fig.~\ref{fig:dmf_others} (right) (all scaled to same cosmology and $\kappa$), we also recalculate the literature values using the SSF fit parameters from Table \ref{tab:schechterTab}, this ensures that they are integrated down to our mass limit. Our measurement is consistent with \citet{Dunne2011},  \citet{Vlahakis2005}, \citet{Clemens2013} and with the lower range of \citet{Driver2007} but smaller than the \cite{Clark2015} values.  However, the latter measurement is subject to a large uncertainty due to cosmic variance  (46.6\,per\,cent, \citealp{Driver2007}) in comparison to the 7-17\,per\,cent for this work\footnote{The cosmic variance in the \citet{Dunne2011} study is 25.7\,per\,cent for $z<0.1$ \citep{Driver2011}.}. Further discussion on the evolution of the dust properties over cosmic time is provided in \citet{Driver2017}.

\subsection{Eddington Bias in the Dust Mass Function}
\label{sec:deconvolve}

{ Here we check whether our DMF is biased due to the dust mass errors from {\sc magphys}.  Since the scatter due to the mass error could move galaxies into neighbouring bins in either direction, and as the volume density is not uniform, this could have the effect of introducing an Eddington bias \citep{Eddington1913} into the DMF.  \cite{Loveday1992} showed that this bias effectively convolves the underlying DMF with a Gaussian with width equal to the size of the scatter in the variable of interest (here dust mass) to give the observed DMF. This is valid assuming that the parameter uncertainties, and hence resulting errors, have a Gaussian distribution.  Here we test whether we can correct for the Eddington bias in the DMF by deconvolving our observed DMF and attempt to extract the underlying `true' DMF. { We expect that any bias in the overall cosmic dust density will be small since galaxies with at least one measurement over 3$\,\sigma$ in one of the {\it Herschel} SPIRE bands contribute around four times as much to the dust density of the Universe than those without a 3$\,\sigma$ measurement in the FIR regime. }

We fit a SSF convolved with a Gaussian, where we estimate the width of the Gaussian using two methods.  First, we derive the width of the convolved function by calculating the mean dust mass error from {\sc magphys} as a function of mass (the varying error method). Second, we take the mean value of the error in dust mass around the knee of the single Schechter function where the convolution will have the strongest effect (where the mean error is $0.11\,\rm dex$, the constant error method). Both produce very similar deconvolved Schechter function fit parameters that are in agreement with the traditional Schechter function method within a few per\,cent.  The deconvolved fit parameters derived with constant error are listed in Table~\ref{tab:schechterTab}; this produces a dust mass density of $(\pVmaxConOmegad) \times 10^{-6}$ for both the $pV_{\rm max}$ and BBD DMFs.  We find that the traditional single Schechter function is a better fit ($\Delta \chi^2 \sim 0.75$) than the deconvolved constant error function, and the varying error method produces a comparable goodness of fit to the traditional SSF without deconvolution.  The reason that the best fit is insensitive to the mass errors is that the mass errors are a strong function of mass: for low mass galaxies, the errors are large ($\sim$ 0.5\,dex); while for higher masses ($\sim \thinspace M^*$), the errors are small ($<$0.1\,dex). At low masses the DMF is a power law, the slope of which is unchanged when convolved by a Gaussian. At higher masses, near the exponential cut-off, the errors are small, and so the effect on the knee is negligible. We therefore conclude that there is no strong argument for choosing to use the deconvolved SSF fits instead of the original single Schechter functions, therefore we include the results here for completeness but continue using the original SSF fits throughout the paper.
}

\begin{figure}
 \includegraphics[width=\columnwidth]{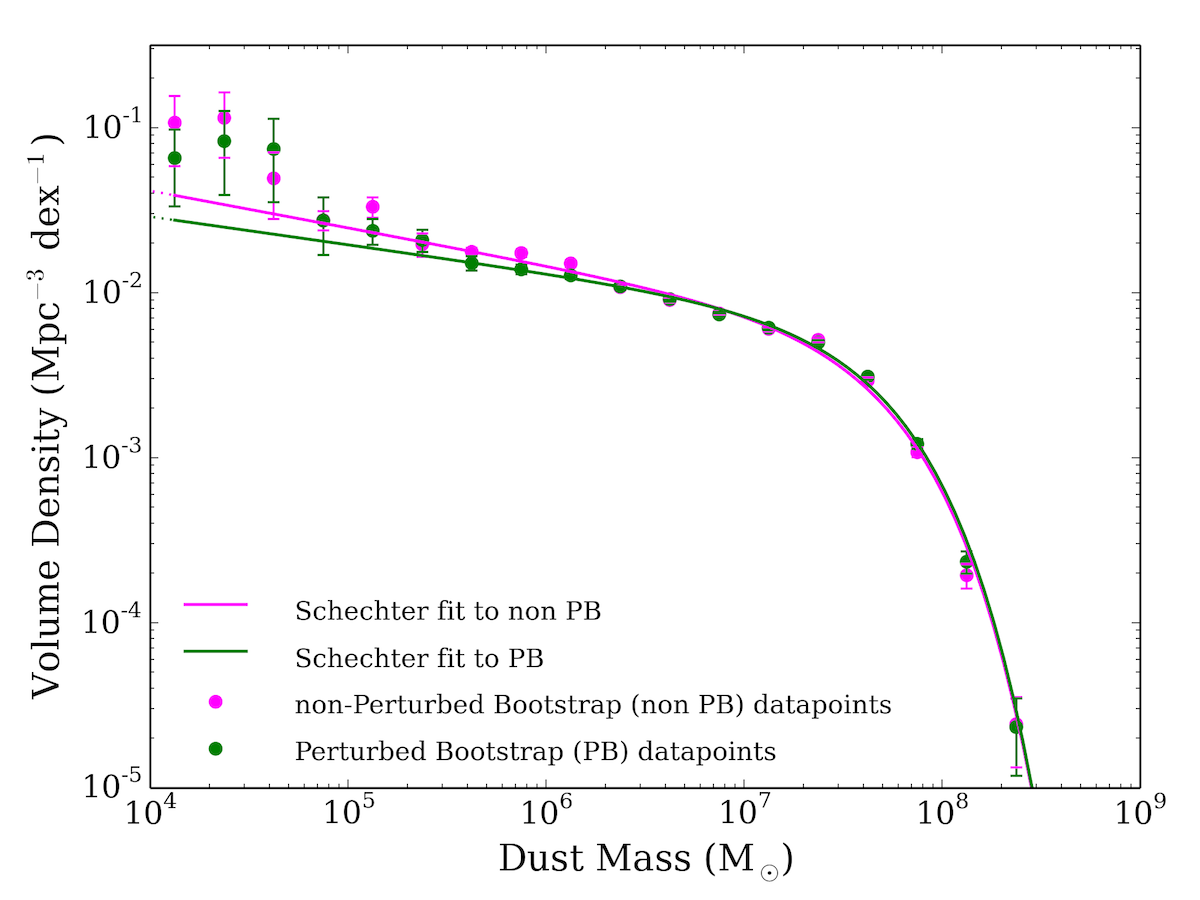}
  \includegraphics[width=\columnwidth]{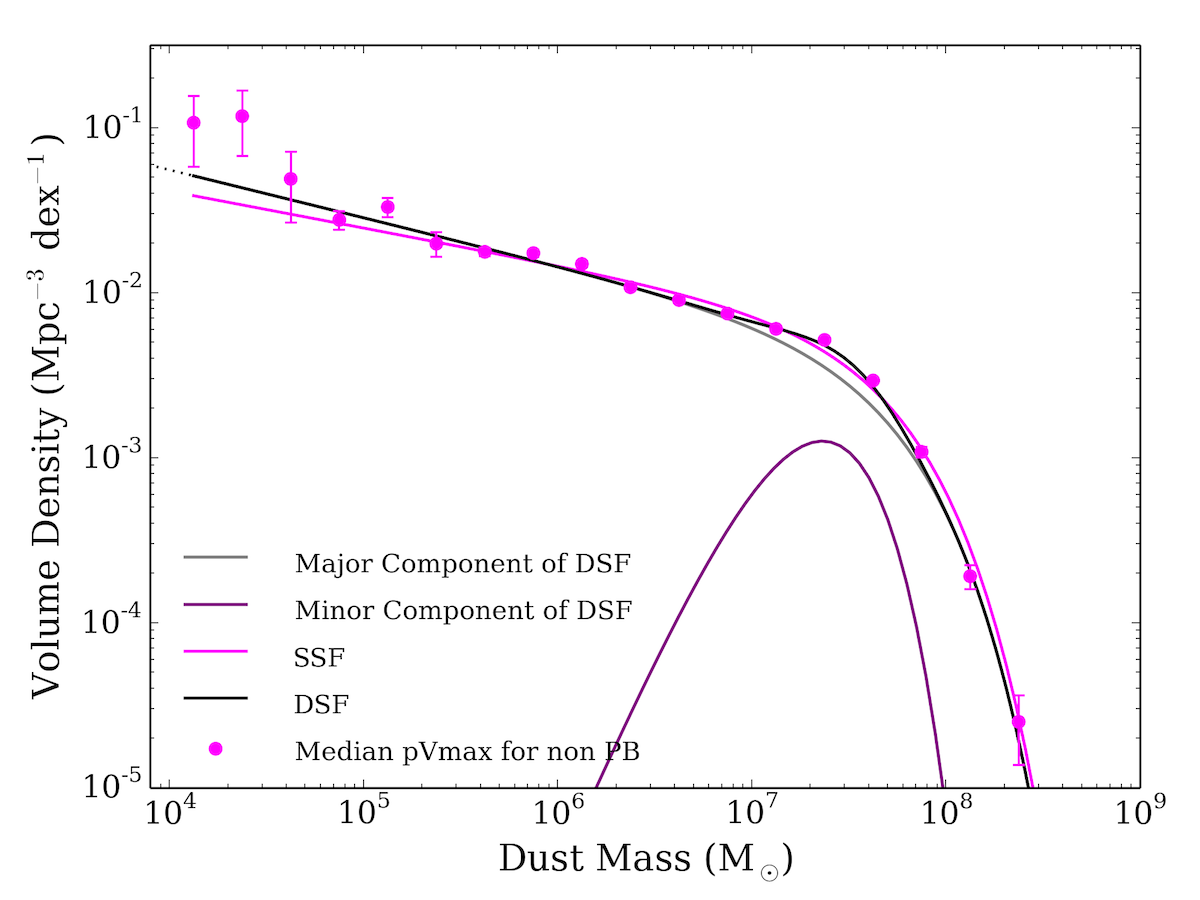}
 \caption{ {\it Top:} The $pV_{\rm max}$ DMF (purple) from the SB measurements compared to the  DMF derived using the bootstrap perturbed (PB) by the uncertainties in the dust mass estimates from {\sc magphys} (the PB DMF, green).   The data points show the volume densities in each mass bin and the solid lines are the best fitting ($\chi^2$) single Schechter functions, SSF, to the data.  {\it Bottom:} Comparison of the SSF with the DSF including the major and minor components. The data points show the volume densities in each mass bin, the major and minor components are shown in grey and purple respectively, the overall DSF is shown in magenta. Error bars are derived from a bootstrap analysis and the data points have been corrected for over and under densities in the GAMA fields (see W17). The total number of sources in each bin is shown in the top panel. }
 \label{fig:pbdmf}
\end{figure}

{
In principle, the difference between the DMFs simple bootstrap method (the SB) and the bootstrap method where we perturbed the data by the underlying uncertainties in dust mass (the PB) provides us another method to test whether the DMF is biased and could provide a way to correct for this.  Fig.~\ref{fig:pbdmf} compares the data and resulting Schechter fits for the observed $pV_{\rm max}$ DMF (SB DMF) and the perturbed (including the uncertainties in the dust mass from {\sc magphys}) $pV_{\rm max}$ DMF (PB DMF). We see that the two DMFs are very similar with fit properties differing by only a few per\,cent. The largest differences in the DMF are seen at the noisier low dust mass end suggesting the biases are indeed small, we believe this is because the uncertainties in the DMF around the knee are small. 

{ We also perform another test to quantify the bias introduced to the DMF by the inclusion of sources with poor FIR constraints. We use the distribution of temperatures of sources with high total FIR signal to noise to define a new temperature prior. Then for each bootstrap sample we draw new temperatures from this prior and adjust the dust masses accordingly. In this way we perform another kind of perturbed boostrap in which each realisation has a temperature distribution that matches the high signal to noise galaxies. We find that the bias introduced to the DMF in this way is very small, and so we believe our DMF is robust. This is discussed in more detail in Appendix \ref{Appendix:Temperature}. }

}


\subsection{A Double Schechter Fit to the DMF}
\label{sec:double}

{
The issues revealed in Section \ref{sec:otherdmfs} (and in Appendix~\ref{Appendix:mmin}) showing the dependence of the SSF fit parameters with the chosen lower mass limit of the SSF fit suggests that the observed DMF is not adequately represented by the SSF. W17 also found that a SSF fit was not sufficient to fit their stellar mass function of the same sample, instead they required a double Schechter function (DSF) fit with the same $M^*$, but different faint-end slopes.  We therefore follow W17 and fit a DSF $D(M)$ but unlike W17, we do not couple the two $M^*$ values, since there is no reason to believe that multiple populations in the dust mass functions would have the same characteristic mass.  The DSF is therefore just defined as the sum of two single functions) of the form:

\begin{align}
  D (M; M^*_1, M^*_2, \alpha_1, \alpha_2, \phi^*, f_{\rm mix}) & = S(M; M^*_1, \alpha_1, \phi^*) \times f_{\rm mix} \nonumber \\
  & + S(M; M^*_2, \alpha_2, \phi^*) \times \left(1-f_{\rm mix}\right)
  \label{eq:double}
\end{align}

\noindent where $f_{\rm mix}$ is the fractional contribution of one of the components. Fig.~\ref{fig:pbdmf} compares the DSF with the SSF. The major component of the DSF is similar to the SSF, but the former provides a better fit to the `shoulder' in the data at $M\sim 10^7\,M_{\odot}$ and results in a reduced $\chi^2 \sim 3 \times$ lower than the SSF fit. Although the DSF significantly reduces the $\chi^2$ of the best fit, the variation of mass errors as a function of mass could introduce this kind of shape in the DMF. We therefore cannot be sure that the DSF represents a fundamentally better model of the data, and prefer to use the SSF as our standard fit. The best-fit parameters for the DSF are listed in Table~\ref{tab:schechterTab}. The dust density for the $pV_{\rm max}$ DSF fit is $\pVmaxDSFOmegad \times 10^{-6}$, corresponding to an overall fraction of baryons (by mass) stored in dust $f_{m_b} = (2.51 \pm 0.04) \times 10^{-5}$, assuming the Planck $\Omega_{\rm b} = 45.51 \times 10^{-3}h_{70}^{-2}$ \citep{Planck2016cosmo}. The DSF therefore returns exactly the same value for dust density as using the simpler SSF. { We also note that the improvement in $\chi^2$ from SSF to DSF becomes insignificant when the uncertainty due to cosmic variance is included in the fitting process.

It is tempting to link the two Schechter components to early and late-type galaxies, but the parameters of the minor component of the DSF do not match those of the early-types (see Section~\ref{sec:split_DMF}). This suggests that the two components of the DSF do not represent physically distinct populations, and so does not provide a better representation of the data.
}

}


\section{Theoretical predictions from galaxy formation models}
\label{sec:theory}

\begin{figure}
 \includegraphics[width=\columnwidth]{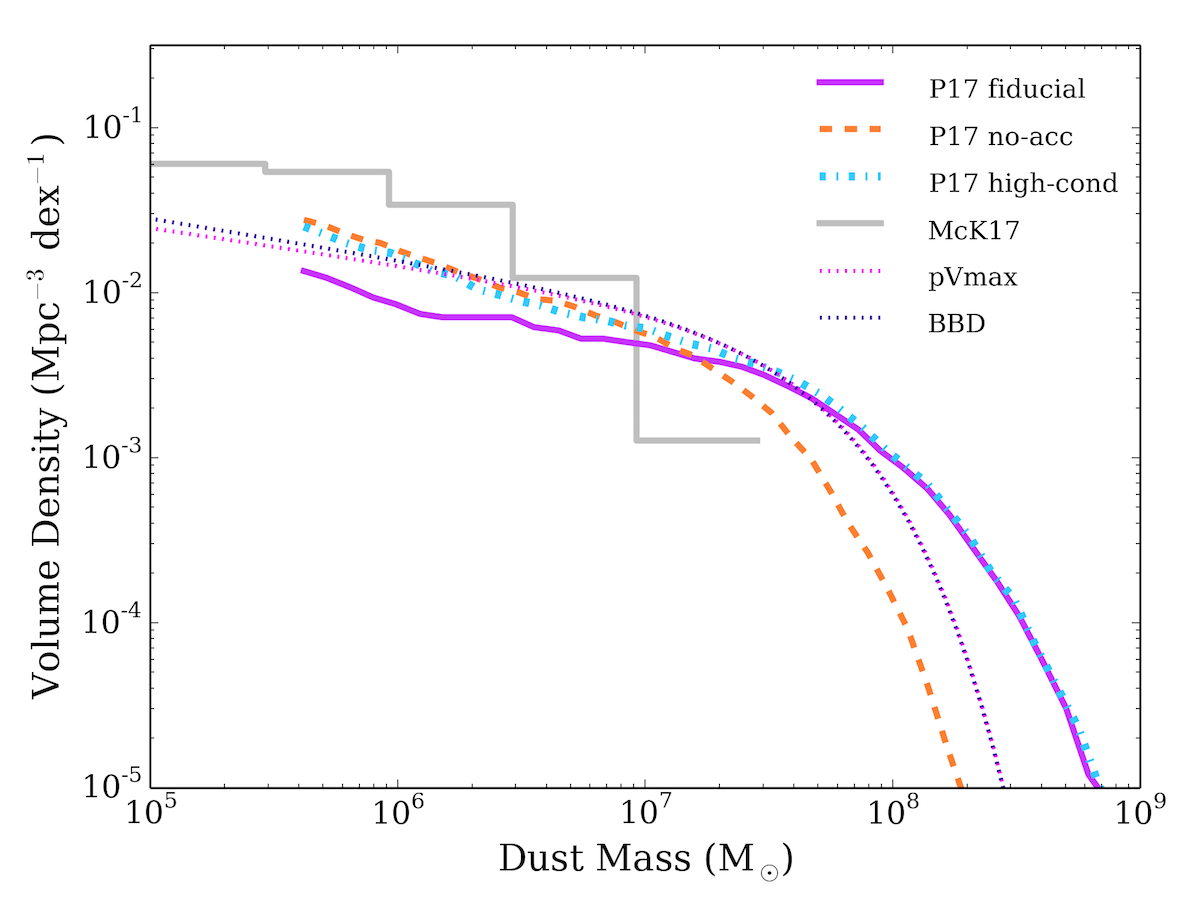}
 \includegraphics[width=\columnwidth]{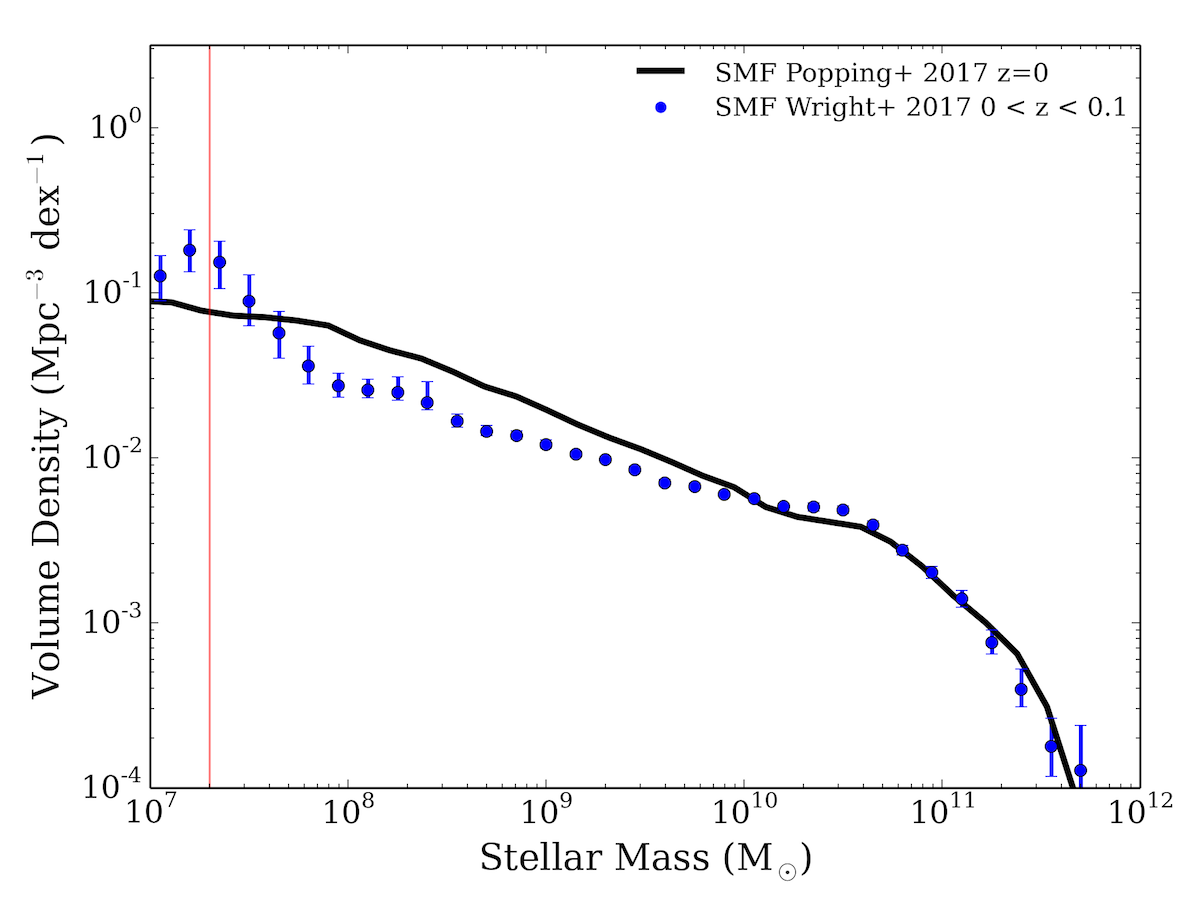}
 \caption{{\it Top:}  A comparison with the predicted $z=0$ DMFs from \citet{Popping2017} (P17) and \citet{McKinnon2017} (McK17) with the { SSF fits} derived from the BBD and $pV_{\rm max}$ methods, see also Table~\ref{tab:popping}. We include three models from P17: the fiducial, no-acc and high-cond models which consist of varying dust condensation efficiencies in stellar winds, supernovae and grain growth in the interstellar medium respectively.  The McK17 histogram is their L25n512 simulation at $z=0$ (their Fig.~2). {\it Bottom:} Comparing the $z=0$ stellar mass functions for the GAMA sources (W17, in blue) with that derived using the SAMs of \citet{Popping2017} (in black). W17 is the SMF of the same optical sample from which our DMF is derived. The vertical line shows the boundary at which W17 fit their data with a Schechter function. }
 \label{fig:popping}
\end{figure}

{ Next we compare the SSF fit to the DMF} from Section~\ref{sec:deconvolve} with theoretical predictions for $z=0$ from the dust models of \citet{Popping2017} and \citet{McKinnon2017}.  \citet{Popping2017} derive DMFs from semi-analytic models (SAMs) of galaxy formation based on cosmological merger trees from \citet{Somerville2015} and \citet{Popping2014} and include prescriptions for metal and dust formation based on chemical evolution models.  They predict DMFs at different redshifts using dust models with dust sources from stars in stellar winds and supernovae (SNe), grain growth in the interstellar medium, and dust destruction by SN shocks and hot halo gas (see also \citealt{Dwek1998,Morgan2003,Michalowski2010,Dunne2011,Asano2013,Rowlands2014,Feldmann2015,DeVis2017b}). Note that for consistency, we have scaled the Popping DMFs down by a factor of 2.39 in dust mass since their $z=0$ models were calibrated on dust masses for local galaxy samples from the \textit{Herschel Reference Survey} \citep{Boselli2010,MWLSmith2012,Ciesla2012} and KINGFISH \citep{Skibba2011} where \citet{Draine2003} dust absorption coefficients are assumed. After this scaling, their DMF (based on their SAMs) is consistent with a Schechter function with $M^* \sim 10^{7.9}\, \rm M_{\odot}$.   In Fig.~\ref{fig:popping} (top) we compare three of their $z=0$ DMF models as defined in Table~\ref{tab:popping}: the so-called fiducial, high-cond and no-acc models. Their fiducial model assumes 20\,per\,cent of metals from stellar winds of low-intermediate mass stars (LIMS) and SNe are condensed into dust grains, with interstellar grain growth also allowed. The high-cond assumes that almost all metals available to form dust that are ejected by stars and SNe are condensed into dust grains, with additional interstellar grain growth. The no-acc model assumes 100\,per\,cent of all metals available to form dust that are ejected by stars and SNe are condensed into dust grains, with no grain growth in the ISM.

\begin{table*}
    \begin{tabular}{lccccccc}
    \hline
    Model Name & \multicolumn{2}{c}{Efficiency dust LIMS} & \multicolumn{2}{c}{ Efficiency dust Type I/II SNe}  & grain growth$^a$ & \multicolumn{2}{c}{dust destruction$^b$}  \\ \hline
             & Carbon & other $Z$ & Carbon & other $Z$ & & SNe & halo \\
						              & (not in CO)& (Mg,Si,S,Ca,Ti,Fe) & (not in CO) & (Mg,Si,S,Ca,Ti,Fe) & & &  \\ \hline
  Popping & & & & & & & \\
  fiducial          & 0.2 & 0.2 & 0.15 &  0.15 & Y, $t_{\rm acc,0}=15$\,Myr & Y & Y \\
	high-cond         & 1.0 & 0.8 & 1.0 & 0.8 & Y, $t_{\rm acc,0}=15$\,Myr & Y&  Y\\
	no-acc       & 1.0 & 1.0 & 1.0 & 1.0 & N & Y&  Y \\ \hline
  McKinnon & & & & & & & \\
  McK16        & 1.0 & 0.8 & 0.5 & 0.8 & Y, fixed $t_{\rm acc}=200$\,Myr & Y & N \\
	McK17        & 1.0 & 0.8 & 0.5 & 0.8 & Y, $t_{\rm acc,0}=40$\,Myr & Y &  Y \\ \hline
    \end{tabular}
    \caption{The dust models used in cosmological predictions of the DMF including three models from \citet{Popping2017} and two models from \citet{McKinnon2016,McKinnon2017}. All of the models presume dust formation in LIMS (low-intermediate stars) in their stellar wind AGB phase and in Type Ia and II supernovae. $^a$ - the timescale for interstellar grain growth in Milky Way molecular clouds such that the grain growth timescale of the system $t_{\rm acc}$ is either fixed or derived from $t_{\rm acc} \propto t_{\rm acc,0} n_{\rm mol}^{-1} Z^{-1} $ where $Z$ is the metallicity and $n_{\rm mol}$ is the molecular number density. $^b$ - destruction of dust by either SN shocks in the warm diffuse ISM or via thermal sputtering in the hot halo gas.  In \citet{Popping2017} 600\,$M_{\odot}$ and 980\,$M_{\odot}$ of carbon and silicate dust are assumed to be cleared by each SN event respectively. In \citet{McKinnon2017} dust destruction is derived in each cell of the simulation, with each SN releasing $10^{51}\,\rm ergs$; this is consistent with their shocks clearing out 6800\,$M_{\odot}$ of gas.  }
    \label{tab:popping}
\end{table*}

The fiducial and high-cond models overpredict the number density of galaxies in the high dust mass regime, $>10^{7.5}\,M_{\odot}$. The no-acc model is the closest model to the observed high mass regime of the DMF, though underestimates the volume density around $M^*$ compared to our DMF (dotted lines in Fig. \ref{fig:popping}).  Both the no-acc and high-cond models are better matches at low masses ($<10^{7}\,M_{\odot}$), while the fiducial model underpredicts the volume density in this regime. This likely suggests that LIMS and SNe have to be more efficient than the fiducial model at producing dust in low dust-mass systems i.e. the dust condensation efficiencies in both stellar sources need to be larger than 0.3, or that the dust destruction and dust grain growth timescales in the fiducial model need to be increased and decreased respectively.  At high masses, the fiducial and high-cond models appear to be forming too much dust. This implies that dust production and destruction are not realistically balanced in these models. This is likely due to the model introducing too much interstellar gas and metals, which allow for very high levels of grain growth in the ISM.

We note that the no-acc P17 model (without grain growth in the ISM) is likely not a valid model as it assumes 100\,per\,cent efficiency for the available metals condensing into dust in LIMS and SNe which is unphysically high, see e.g. \citet{Morgan2003,Rowlands2014}. { Hereafter we no longer discuss this model even though by eye it appears to be an adequate fit to the observed DMF at masses below $\rm 10^{7}\,M_{\odot}$.}

\begin{figure*}
 \includegraphics[trim=0mm 6mm 10mm 5mm clip=true,width=0.95\textwidth]{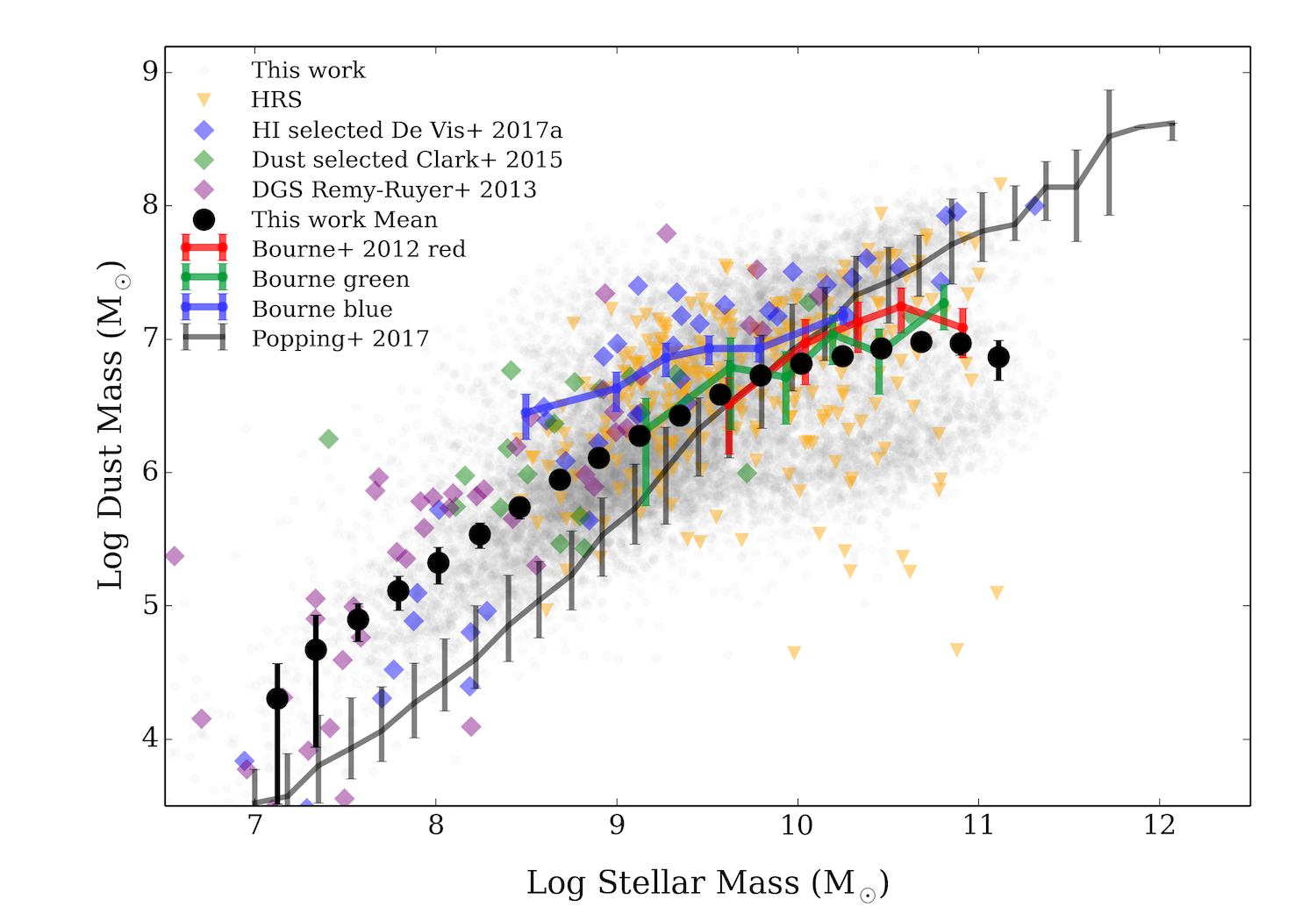}
 \caption{The dust to stellar mass ratio for galaxies in the local Universe. { The data from this work is shown in the underlying grey points with mean dust masses ($\pm$ standard error) in each stellar mass bin (black).  We include a compilation of \herschel\ results for local galaxies including the stellar-mass selected HRS \citep{Boselli2010}, the dust-selected sample of \citet{Clark2015}, the H{\sc i}-selected sample from \citet{DeVis2017a} and the dwarf galaxy survey from \citet{Remy-Ruyer2013}.  Overlaid are the local universe relationships ($z<0.12$) based on stacking on \herschel~maps for 80,000 galaxies from \citet{Bourne2012} in three different $g-r$ colour bins (their Fig.~16). All of these samples have been scaled to the same the $\kappa$ value with parameters derived using {\sc magphys} (see \citealt{DeVis2017a}) or modified blackbody fitting but scaled to the same $\kappa$ (DGS and stacked samples).   The median dust and stellar masses from \citet{Popping2017} are shown by the grey line with 16 and 84\,percentile errorbars (scaled by a factor of 1/2.39 in dust mass).   }}
 \label{fig:poppingduststars}
\end{figure*}

To investigate the discrepancy between the observed DMF in this work and the predicted SAM DMF from \citet{Popping2017}, we first check that the stellar mass function from the SAMs is consistent with the observed galaxy stellar mass function (GSMF) for the GAMA sample in W17 (Fig.~\ref{fig:popping} bottom). The SMFs at the high mass end are in agreement though the model SMF has a slight overdensity of galaxies in the range $10^8<M_{\rm s} \, (M_{\odot})<10^{9.4}$, where $M_{\rm s}$ is stellar mass.  If this overdensity of sources were responsible for the discrepancy between the predicted and observed DMFs in the high $M_{\rm d}$ regime, those intermediate stellar mass sources would have to have dust-to-stellar mass ratios of $\sim 0.5$ which is again unphysical.  We can see this is not the case when comparing the dust-to-stellar mass ratios of the \citet{Popping2017} fiducial $z=0$ model in Fig.~\ref{fig:poppingduststars} (as mentioned earlier, this is based on \herschel\ observations of local samples of galaxies).


In Fig.~\ref{fig:poppingduststars} we plot the dust and stellar masses from the compilation of local galaxy samples collated in \citet{DeVis2017a,DeVis2017b}\footnote{these have all been scaled to the same value of $\kappa$ and apart from the Dwarf Galaxy Survey, all galaxy parameters have been derived using the same fitting techniques.} and compare with P17 and our sample of $\sim$15,000 sources. Here we can clearly see the cause for the discrepancy between the observed DMF from this work and the model: \emph{the model overpredicts the amount of dust in high stellar mass sources, well above any dust-to-stellar ratios observed locally}. Although the observations show a flattening of dust mass at the highest $M_{\rm s}$ regime (where early type galaxies are dominating), this is not the case in the SAM.   In general the SAM prediction assumes a constant dust-to-stars ratio of $\sim 0.001$ across all mass ranges. The observations however suggest that there is a roughly linear relationship until $M_{\rm s} > 10^{10}\,M_{\odot}$, after which the slope flattens, with $M_{\rm d}/M_{\rm s} \sim 0.001$.

Fig.~\ref{fig:poppingduststars} also suggests that $M_{\rm d}/M_{\rm s}$ increases to $\sim 0.025$ in low stellar mass galaxies (in agreement with \citealt{Santini2014,Clark2015,DeVis2017a}).  This is further supported by the stacking analysis carried out in \citet{Bourne2012} whose dust-to-stellar mass trends in different bins of optical colour are added to Fig.~\ref{fig:poppingduststars}. These were derived by stacking on $\sim$80,000 galaxies in the {\it Herschel} maps, revealing that low stellar mass galaxies had higher dust-to-stellar mass ratios, consistent with these sources having the highest specific star formation rates. Our binned data (black points) are in agreement with local galaxy surveys and the \citet{Bourne2012} trends: we see that the slope of dust-to-stellar mass flattens at high masses, and that there exists a population of dusty low-stellar-mass sources that the SAM does not predict. 

Alternative predictions for a local DMF are provided by \citet[hereafter McK16, McK17]{McKinnon2016,McKinnon2017}. In these models, dust is tracked in a hydrodynamical cosmological simulation with limited volume. The McK16 dust model is similar to the P17 high-cond model (including interstellar grain growth and dust contributed by both low mass stellar winds and SNe) but has no thermal sputtering component. The updated model from McK17 reduces the efficiency of interstellar grain growth and includes thermal sputtering (see Table~\ref{tab:popping}).  The DMF from McK17 (their L25n512 simulation at $z=0$) is shown in Fig.~\ref{fig:popping} (top).  Their values have been scaled to the same cosmology as used here (they use the same $\kappa$ and Chabrier IMF as this work).  We can see that McK17 predicts fewer massively dusty galaxies than P17 and our observed DMFs.
Although their DMF fails to produce enough galaxies in the highest mass bins in Fig.~\ref{fig:popping}, the simulated DMF becomes more strongly affected by Poissonion statistics in this regime due to the small volume of the simulation.

Possible explanations for the difference between the predicted (P17, Mck17) and observed DMFs at large dust masses are (i) the efficiency of thermal sputtering due to hot gas in the halo has been under or overestimated in these highest stellar mass sources; (ii) the fiducial and high-cond dust models of P17 allow too much interstellar grain growth in highest stellar mass galaxies due to the assumed timescales or efficiencies of grain growth being too high; (iii) the predicted highest stellar mass galaxies have too little (McK17) or too much (P17) gas reservoir potentially due to feedback prescriptions being too strong/not strong enough, respectively. If the gas reservoir is too high, interstellar metals can continue to accrete onto dust grains and increase the dust mass. Conversely if it is too low, then the contribution to the dust mass via grain growth will be reduced.  We will adress each of these possibilities in turn.

(i) We can test if the amount of dust destruction by thermal sputtering in hot (X-ray emitting) gas could explain the differences in the predicted and observed DMF at the high mass end as McK16 and McK17 already compared the results using dust models without and with thermal sputtering respectively. They find that including thermal sputtering only makes small changes to the shape of DMF since this affects dust in the halo rather the interstellar medium, this is therefore not likely to be responsible for the disreprancy.

(ii) Comparing the dust models in P17, McK16 and McK17 allow us to test the effect of changing the grain growth parameters.   The timescale for grain growth is shortest in P17 and McK16 and both those models produce too much dust in the high dust mass regime of the DMF.  McK17 has a longer grain growth timescale ($t_{\rm acc,0} =40\,\rm Myr$, Table~\ref{tab:popping}) than both P17 and McK16 and this change indeed reduces the volume density of the highest dust mass sources.   McK17 also compares the DMFs from the same simulation methods with different dust models and they find that a significantly reduced DMF at the high mass end can be attributed to the longer grain growth timescales.

(iii) Earlier we showed that the galaxies that are responsible for the highest dust mass bins in the P17 DMF have too much dust for their stellar mass (Fig.~\ref{fig:poppingduststars}). To test whether they have too much dust due to the gas reservoir of the SAM massive galaxies being too high (hence leading to more interstellar grain growth) we refer to the predicted and observed gas mass function comparison in \citet{Popping2014}. There they showed that these are not as discrepant as we see here with the modeled and observed dust mass functions and therefore are likely not responsible for the discrepancy in the DMF.

We therefore conclude that it is likely that the interstellar grain growth in these massive galaxies is simply too efficient/fast in the P17 and McK16 dust models. In this scenario, the few largest stellar mass galaxies are allowed to form too much dust in the interstellar medium at a rate that is not observed in real galaxies. However, the growth timescale may also be too slow in the McK17 model.  All of the P17 high-cond, McK16 and McK17 dust models assume very high dust condensation efficiencies in AGB stars and Type Ia and II SNe.  We propose therefore that the most realistic dust model must lie somewhere in between these and the fiducial P17 model, with stardust condensation efficiencies larger than 0.3 but lower than 0.8 and a similar dust grain growth timescale as assumed in P17.

Neither the P17 fiducial, nor the McK16 and McK17 dust models provide reasonable matches to the low dust mass regime ($\sim 10^{7.5}M_{\odot}$) of the DMF.  McK16 and McK17 overpredicts the dust masses in the low mass regime and P17 fiducial model underpredicts the DMF suggesting again that stardust condensation efficiences may be intermediate between the three models with a grain growth timescale similar to P17.  Only the P17 high-cond model provides an adequate match to this regime. Based on the observed DMFs, we will explore ways to improve the theoretical models in future work.

\section{The DMF by morphological type}
\label{sec:split_DMF}

{ We can test the standard prejudice that spiral galaxies are full of dust, and ellipticals have very little dust by using the DMF to quantify the difference in dust content of early-type galaxies (ETGs) and late-type galaxies (LTGs)}  We create ETG and LTG subsets for our sample of galaxies based on classifications carried out by GAMA in \citet[hereafter D12]{Driver2012} and \citet[hereafter M16a]{Moffett2016_gals}. For both studies visual classifications were based on three colour images built from H (VIKING), $i$, and $g$ (SDSS) bands.  Classifications were based on three pairs of classifiers in which there was an initial classifier and a classification reviewer.  D12 classified the entire sample out to $z \le 0.1$ and split the sample only into `Elliptical' and `Not Elliptical' galaxies, which we hereafter refer to as ETGs and LTGs (later type galaxies).   The classifications from M16b were carried out on the same sample as D12, but limited to $z \le 0.06$. In M16a they attempted to produce an updated set of morphological classifications using classification trees with a finer binning system than D12.  However, for consistency with the D12 classifications (and because here we do not want to split the DMF into finer morphological classes), we include the M16a Ellipticals in the ETG class, and we group all remaining galaxy types apart from lttle blue spheroids (LBSs) into the LTG category (following M16a we keep the LBS sources separate). We note that if LBSs are included in the LTG catagory there is very little change in either $M^*$ or $\phi^*$; however, the low-mass slope is steepened by $\sim 6$\,per\,cent, overall the difference in the dust mass density made by including LBSs in the LTG catagory is only around 2\,per\,cent.


We next use these morphologies in order to investigate the shape and dust mass density of ETGs and LTGs. We choose to limit our redshift range to $z \le 0.06$ for two reasons (i) the finer, updated classifications of D12 provided in M16a is limited to this range and beyond this range visual classifications become more uncertain (ii) with increasing redshift, the sample will suffer more from incompleteness at lower masses. This is demonstrated in Fig.~\ref{fig:dust_stellar_binned} where we compare the dust and stellar masses of the ETGs and LTGs from D12 at $z \le 0.06$ and $0.06 < z \le 0.1$. We see a dearth of galaxies below $M_{\rm d}\sim10^{5.5}$ and stellar masses below $M_{\rm s}\sim10^{7.5}$ in the higher redshift bin.

\begin{figure}
 \includegraphics[width=\columnwidth]{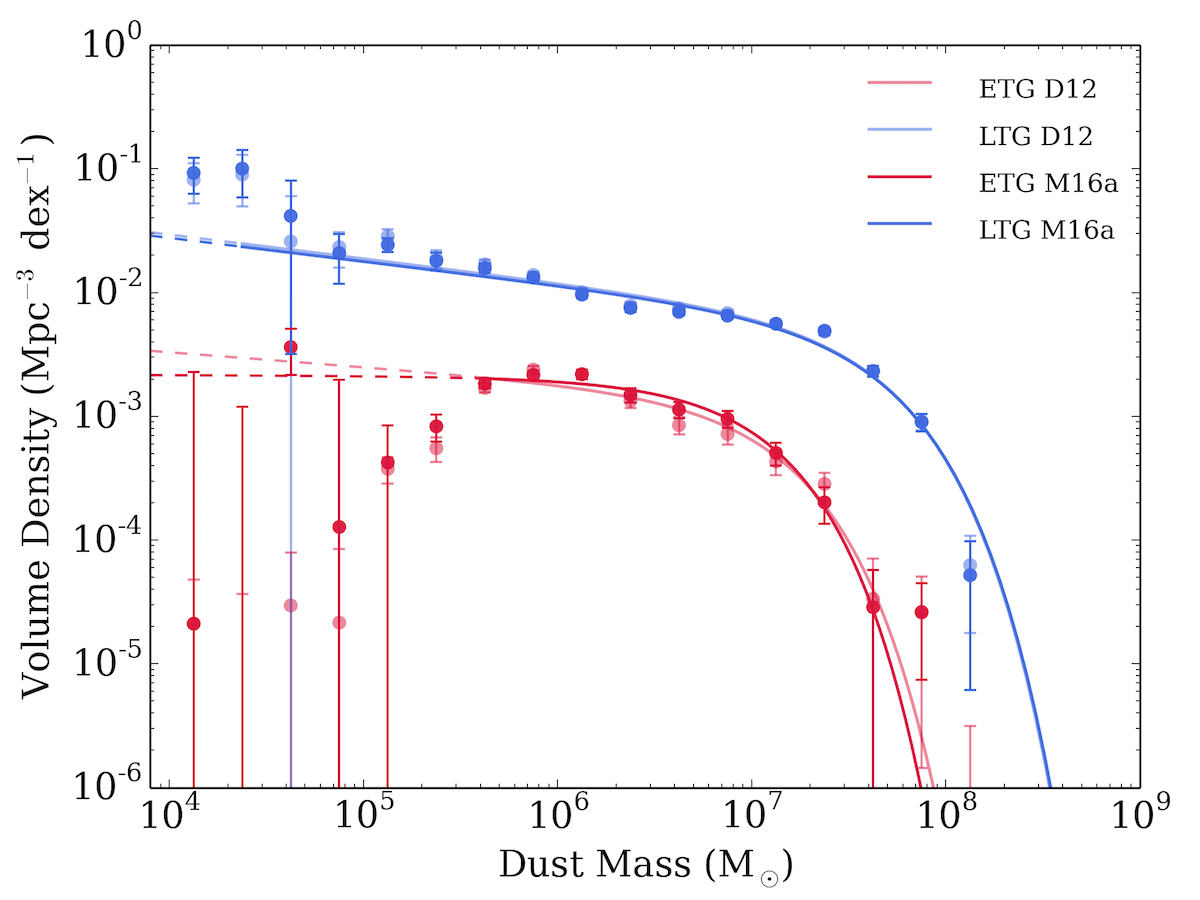}
 \caption{ The $pV_{\rm max}$ dust mass functions for the GAMA/{\it H-}ATLAS sources at $0.002<z<0.06$. Here the opaque lines show the sample for the fitted range split into ETGs and LTGs by \citet{Moffett2016_gals} - M16a, and the translucent lines show the sample for the fitted range as split into ETG and LTG by \citet{Driver2012} - D12.  Red denotes ETGs and blue the LTGs.  The data points show the observed values and the solid lines are the best fitting ($\chi^2$) single Schechter functions to the data for their respective fitted regions, beyond this we show extrapolations down to $10^{4} M_{\odot}$ as dashed lines. Error bars are derived from a bootstrap analysis and the data points have been corrected for over and under densities in the GAMA fields (see W17). }
 \label{fig:dmf_EnotE}
\end{figure}

In our final $z<0.06$ sample, a total of 5736 sources were classified by D12, 588 of which were ETGs, 4837 as LTGs, and 474 as LBS.  In the same redshift range, M16a classified 5765 galaxies with 639 ETGs, 4599 LTGs, and 690 LBSs.  There are 773 disagreements between the two sets of classifications (13\% of the overall sample).  The resulting DMFs from the two classification methods are displayed in Fig.~\ref{fig:dmf_EnotE} ({ using SSF fits}), the red and blue points are ETG and LTG respectively, and translucent and solid represent the DMFs for the early and late populations as defined by D12 and M16a respectively. The DMFs for D12 and M16a agree well out to $z\le 0.06$ showing that although the individual classifications are not an exact match, the shapes of the DMFs of the ETG and LTG populations appear to be consistent between the two different classification methods. The Schechter function fit parameters and dust mass densities for ETGs and LTGs are listed in Table~\ref{tab:schechterTabEvo}. (From now on we choose to discuss only the Schechter fit parameters arising from the M16a classifications as these are the most recent).  Unsurprisingly we see there is an order of magnitude more dust mass contained within LTGs than ETGs at $z\le 0.06$, with dust mass densities of $\Omega_{\rm d} = (\MoffnotElowOmegadinf) \times 10^{-6}$ and  $\Omega_{\rm d} = (\MoffElowOmegadinf) \times 10^{-6}$ respectively. LTGs are mostly responsible for the dust content of the Universe.  However, Fig.~\ref{fig:dmf_EnotE} demonstrates that the ETG DMF is not well described by a Schechter function, indeed there is a significant downturn in the volume density of ETGs below dust masses of 10$^{6}\,M_{\odot}$. We believe this is a real effect since we have no reason to believe that incompleteness may be biasing our measurements at this redshift range. The downturn is also in line with the shape of the galaxy stellar mass function (GSMF) for this subset as measured by M16a. Since the DMF for the ETGs clearly does not match a Schechter function, the dust mass density in Table~\ref{tab:schechterTabEvo} may be overestimated. We therefore also calculate the dust mass density for these galaxies by summing the contribution from each galaxy and derive a density of $\Omega_{\rm d} = (\MoffElowOmegadsum) \times 10^{-6}$. This is consistent with the integrated Schechter function since although that fit over-predicts the total dust mass density in low dust mass sources, it also underpredicts the dust mass density at the high mass end. { Comparing our ETG and LTG Schechter fits with the double component fit to the entire sample from Section~\ref{sec:double}, we find that the major component of the double fit matches the high mass end but slightly overshoots the volume densities derived for the LTGs at intermediate masses $10^5< M_{\rm d} (M_{\odot}) < 10^7$ whereas the second component has a peak volume density at higher dust masses than the ETGs.  However, this may be due to misclassification of galaxy types or simply due to the degeneracy in what the fitting routine assigns to each component at the faint end. 
}

\begin{table*}
\centering
    \begin{tabular}{|l|l|c|c|c|c|c|}
    \hline \hline
                                           & Population & $M^{*}$ & $\alpha$ & $\phi^{*}$ & $\Omega_{\rm d}$  & Number of \\
                                           &			  &		($10^7\,h^2_{70}\,\rm M_{\odot}$)  &       &    ($10^{-3}\,h^3_{70}\,\rm Mpc^{-3}\,\rm dex^{-1}$)     &    ($10^{-6}$)  & Galaxies  \\ \hline
\multirow{3}{*}{$0.002<z<0.06$}     & ETG          &   $\MoffElowMstar$    &   $\MoffElowalpha$    &   $\MoffElowphistar$      &    $\MoffElowOmegad$   & 690 \\
                                           & LTG      &   $\MoffnotElowMstar$    &   $\MoffnotElowalpha$    &   $\MoffnotElowphistar$      &    $\MoffnotElowOmegad$   & 4599 \\
                                           & total      &   $\totlowMstar$    &   $\totlowalpha$    &   $\totlowphistar$      &    $\totlowOmegad$   & 5937  \\ \hline \hline
    \end{tabular}
    \caption{ Schechter function fit parameters for the $pV_{\rm max}$ DMFs for the ETG, LTG, and total populations for the low-redshift ($0.002<z<0.06$) subset of our sample using morphological classifications from M16a. }
    \label{tab:schechterTabEvo}
\end{table*}

\section{Comparison with the Galaxy Stellar Mass Function}
\label{sec:GSMF_comp}

\begin{figure}
 \includegraphics[width=\columnwidth]{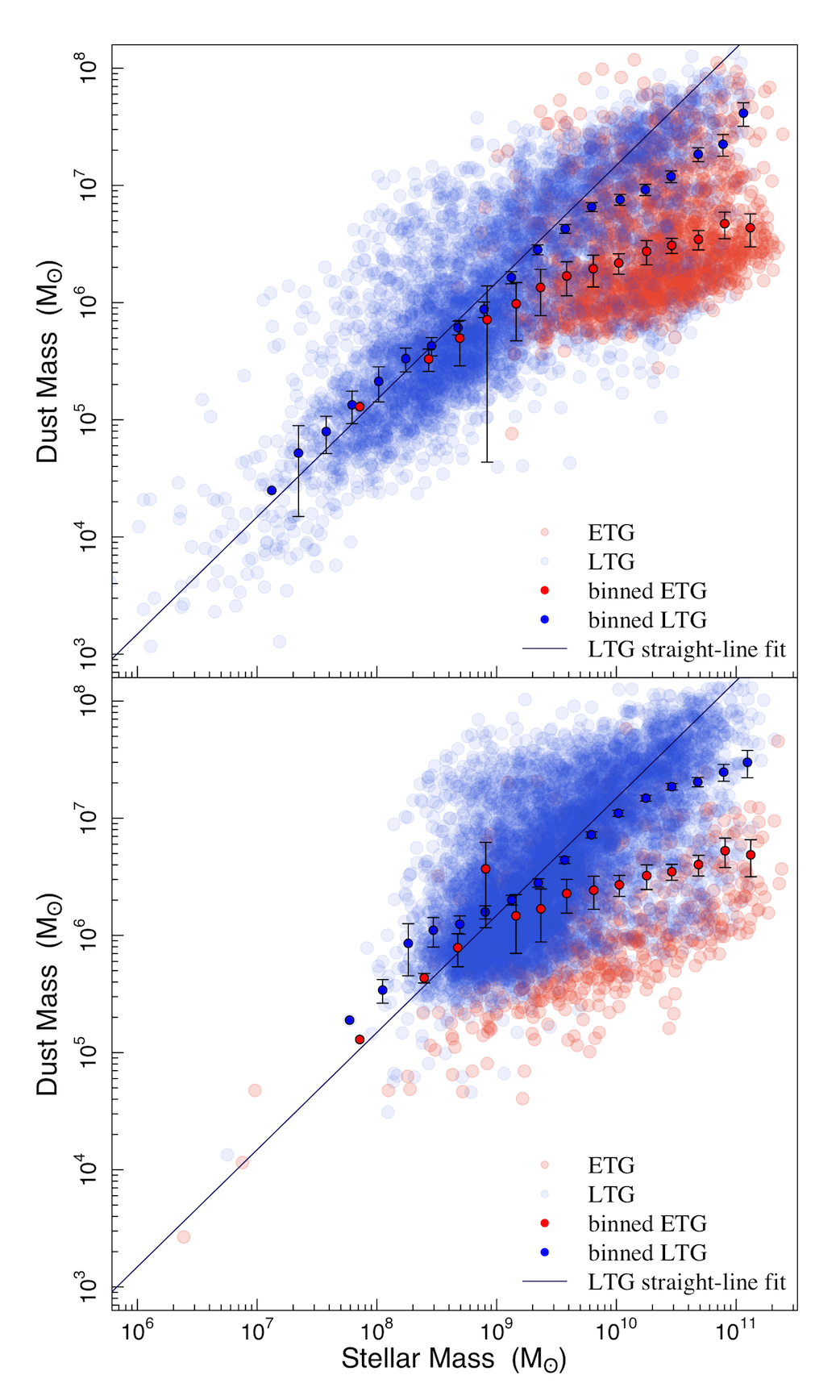}
 \caption{The mean dust to stellar mass ratio for galaxies in our high and low redshift samples ({\it Top:} $0.002 \leq z \leq 0.06$, {\it Bottom:} $0.06\leq z \leq 0.1$) in bins of galaxy stellar mass for ETGs (red), LTGs (blue) in comparison to the ETG and LTG subsamples. We also show in blue a straight-line fit to the LTG data to illustrate the approximate linear scaling of the LTG dust-to-stellar mass ratio at low masses. }
 \label{fig:dust_stellar_binned}
\end{figure}

Scaling relations between dust and stellar mass can reveal the relation between internal galaxy properties and the dust content and whether there is a simple prescription that can tell us how much dust exists in galaxies given a unit of stellar mass (e.g. \citealt{Driver2017}). \cite{Cortese2012} and \cite{MWLSmith2012} investigated $M_d-M_{\rm s}$ scaling relations in local galaxies using the HRS, finding that larger stellar mass galaxies have lower dust-to-stellar mass ratios (see also \citealt{Santini2014}). This was further confirmed in the larger statistical study of \citet{Bourne2012} from H-ATLAS (Fig.~\ref{fig:poppingduststars}).  As we saw in Section~\ref{sec:theory}, the \citet{Popping2017} SAMs produce a trend in $M_d-M_{\rm s}$ which does not agree with the scatter seen in the observations of local galaxies (due to colour, morphological type, environment etc.) and their models produce too much dust in the highest stellar mass disks.  Further modelling of dust and stellar mass scaling relations carried out in \citet{Bekki2013} using chemodynamical simulations reproduced the $M_d-M_{\rm s}$ trend roughly (with massive disk galaxies more likely to have smaller dust-to-stellar mass ratios), but could not reproduce the $M_d-M_{\rm s}$ at stellar masses $>10^{10}\,M_{\odot}$.  In Fig.~\ref{fig:dust_stellar_binned} we bin galaxies in stellar mass, and plot the mean dust mass as a function of mean stellar mass in each bin, with error bars calculated from the standard error on the average. The low-mass end of the LTG dust and stellar mass comparison is actually fairly well-represented by a linear relationship, but diverges at higher masses. { The change in slope at high $M_{\rm s}$ is largely caused by LTGs with $M_{\rm d}$ similar to ETGs at similar $M_{\rm s}$. } It is difficult to disentangle whether this effect is entirely physical due to dust-poor LTGs or due to the misclassification of ETGs as later-type galaxies. It is clear from this plot that the ETGs do not follow a linear trend.

With our dataset we can test whether there is a simple scaling relation between dust and stellar mass by simply comparing the dust and stellar mass functions.   { Since the GSMF in W17 is fit by a coupled DSF with shared $M^*$, it is not equivalent to our DSF, we instead compare the SSF fit} with the GSMFs from M16a who present the GSMFs for ellipticals and for later mophological types including Sab-Scd, SBab-SBcd, Sd-Irr and S0-Sa.
We also compare with \citet[hereafter M16b]{Moffett2016_disks} who have further decomposed the sample into bulges, spheroids and disk components. This decomposition of galaxies into bulge and disk was performed by fitting a double-S\'{e}rsic profile to those galaxies which were morphologically classified as double-component galaxies to obtain bulge-to-total luminosity ratios. From these ratios, the $g-i$ colour and $i$-band absolute magnitudes for both bulge and disk were derived and used, along with the stellar mass relation of \cite{Taylor2011}, to calculate bulge and disk component stellar masses.   
We have already seen that the DMF is dominated by the LTGs in the previous Section, we now test whether most of the dust will be associated with the disk component of galaxies and whether we can scale from stellar mass to dust mass for different galaxy subsamples.  In Table~\ref{tab:Ratio} we compare the ratio of the knee of the Schechter function fit parameters ($M^{*}_{\rm d}/M^{*}_{\rm s}$) for the GSMF and DMF and the integrated mass densities ($\rho_{\rm d}/\rho_{\rm s}$) between different galaxy populations including LTG, ETGs and disks.

\begin{figure}
 \includegraphics[width=\columnwidth]{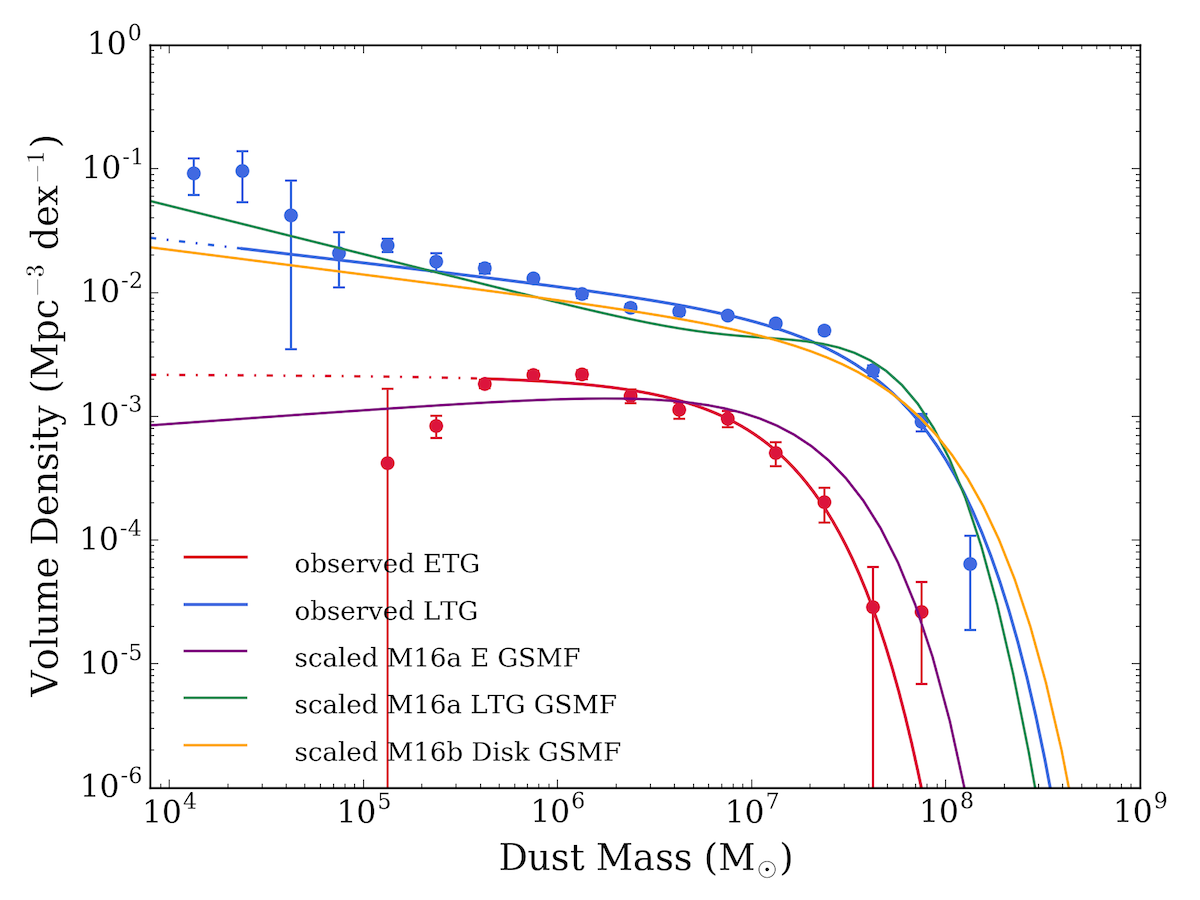}
 \caption{The $pV_{\rm max}$ dust mass functions for the GAMA/{\it H-}ATLAS sources at $0.002<z<0.06$.  The Schechter fits for the ETGs and LTGs are shown by the solid red and blue lines for the fitted range, beyond this we show extrapolations down to $10^{4} M_{\odot}$ as dashed lines. We also compare the GAMA galaxy stellar mass functions from M16a and M16b scaled by $\rho_{\rm d}/\rho_{\rm s}$ including the M16a Elliptical and Late types GSMFs scaled by $\rho_{\rm d, ETG}/\rho_{\rm s, Elliptical}$ and $\rho_{\rm d, LTG}/\rho_{\rm s, (Sab-Scd+Irr+S0-Sa)}$ (purple and green respectively) and the M16b disk GSMF scaled by $\rho_{\rm d, LTG}/\rho_{\rm s, disk}$ (yellow). }
 \label{fig:dmf_EnotE_scaled}
\end{figure}

First we compare the GSMFs to their equivalent DMFs for ETGs and LTGs. We produce a composite Schechter function from the later-type GSMFs from M16a containing the same sample of galaxies as our LTG DMF. We show a version of the LTG GSMF composite Schechter function as scaled by the ratio $\rho_{\rm d}/\rho_{\rm s}$ in Fig.~\ref{fig:dmf_EnotE_scaled}. The scaled LTG GSMF fits the high-mass end of the LTG DMF; however, it diverges from the datapoints around $10^{7} M_{\odot}$ where a more pronounced shoulder is seen in the GSMF than we observe in the DMF. Otherwise  the composite LTG GSMF is in good agreement with our data, and therefore an estimate of the LTG DMF could be made by scaling the LTG GSMF by a factor of $(\rhoLTGBnLTGMt) \times 10^{-4}$.

M16a includes an Elliptical GSMF (equivalent to what we have defined here as ETG), and we scale this function by the $\rho_{\rm d}/\rho_{\rm s}$ in order to compare with the ETG DMF. {  As seen in Fig.~\ref{fig:dmf_EnotE_scaled}, the scaled Elliptical GSMF is not a good match to the ETG DMF data or Schechter function. Compared to the data, the scaled GSMF is too high for $M_{\rm d}>10^{7}$ and $M_{\rm d}<10^{5.5}$, and too low for $10^{5.5} < M_{\rm d}<10^{6.5}$. Indeed, the Elliptical GSMF is also not well-fitted by a Schechter function, and displays the same downturn that we see in the DMF. Whilst the low-mass slope derived by M16a does show a drop-off, it is not as severe as actually observed in either the dust mass or stellar mass function data. } Also, we note that the dust mass as a function of stellar mass for ETGs as seen in Fig.~\ref{fig:dust_stellar_binned} is not consistent with a simple linear relationship.  We therefore caution against using a scaling law between stars and dust that relies upon a Schechter function fit to ETGs. We also note that the ratio $M^{*}_{\rm d}/M^{*}_{\rm s}$ for this sample is $\left(\MstETGBnEMt\right) \times 10^{-4}$, which is of order 17-25 times higher than the average dust-to-stellar mass ratios of Ellipticals compared to recent Herschel studies of ETGs in the local volume ($D<40\rm\,Mpc$, \citealp{SmithMLW2012,Amblard2014}). This may suggest some contamination in our E category from S0s, though we note that the Herschel studies were potentially biased to older redder sources \citep{Rowlands2012, Clark2015} and high density environments \citep{SmithMLW2012,Agius2013}.

Next we attempt to explore the relationship between stellar mass of the disk component and the dust mass of a galaxy. M16b found that the disk GSMF\footnote{We have corrected for the fact that the $\phi^*$ values for the stellar mass functions listed in \cite{Moffett2016_disks} are in units of $\rm Mpc^{-3}$ and not $\rm Mpc^{-3}\,dex^{-1}$.} is well fit by a Schechter function with $\alpha_{\rm disk} = -1.20 \pm 0.02$, which is consistent with our alpha values both for the LTG DMF and the total DMF for the $z<0.06$ sample. Based on the compatibility of the $\alpha$ values, there may be a simple scaling from the disk GMSF to either the total or LTG population DMFs. The scaled function is a good but imperfect fit, and we see a moderate overshoot of the high mass data points by this scaled function, reflecting the fact that the dust-to-stellar mass ratio is not constant. We therefore conclude most of the dust in galaxies is associated with their disk components, and that it is possible to obtain a reasonable representation of the DMF by scaling the disk GSMF by the ratio $\rho_{\rm d}/\rho_{\rm s}=(\rhoLTGBndskMt) \times 10^{-4}$.  We can see that the ratios $M^{*}_{\rm d}/M^{*}_{\rm s}$ and $\rho_{\rm d}/\rho_{\rm s}$ for the disk GSMF and both the total and LTG DMFs shown in Table~\ref{tab:Ratio} are discrepant by more than 1$\,\sigma$. This provides further evidence that the scaling from the disk GSMF to the DMF of either population cannot be exactly linear. We also infer from this that the dust-to-stellar mass ratio is higher for lower-mass disks. We refer the reader to \citet{DeVis2017b} whose work hints that the observed dust-to-stellar mass properties of local galaxies may require the contribution of dust sources from stars and interstellar grain growth to be different for low and high mass galaxies.

Given that the observed dust-to-stellar masses of galaxies are not linear across the whole stellar mass range (Fig. \ref{fig:poppingduststars} and \ref{fig:dust_stellar_binned}), it is somewhat surprising that we can simply scale the GSMFs of LTGs and disks and obtain DMFs close to the observed.  Although the binned dust masses in Fig.~\ref{fig:dust_stellar_binned} at stellar masses greater than $10^{9.5}M_{\odot}$ depart from a linear scaling relation, on average we can assume the slopes are close to linear (especially around the knee of the SF) and therefore this simple scaling appears to work. Surprisingly, this simple scaling from stellar mass to dust mass for LTGs at z=0 (Fig.~\ref{fig:dmf_EnotE_scaled}) may suggest that all the different dust processes (dust condensation in stellar atmospheres and SNe, grain-growth, dust destruction) are correlated to the growth of stellar mass in galaxies. The dispersion in this scale (e.g. Fig.~\ref{fig:poppingduststars} and \ref{fig:dust_stellar_binned}) could therefore place limits on the way dust is formed and how it evolves. We will return to this in future work.

\begin{table}
\centering
    \begin{tabular}{|c|c|c|c|}
    \hline \hline
DMF & Stellar Mass & $M^{*}_{\rm d}/M^{*}_{\rm s}$ & $\rho_{\rm d}/\rho_{\rm s}$ \\
Population & Population & ($10^{-4}$) & ($10^{-4}$) \\ \hline
ETG & Elliptical & $\MstETGBnEMt$ & $\rhoETGBnEMt$ \\ \\
LTG & All Disks & $\MstLTGBndskMt$ & $\rhoLTGBndskMt$ \\ \\
LTG & LTG & - & $\rhoLTGBnLTGMt$ \\ \\
Total & All Disks & $\MsttotBndskMt$ & $\rhototBndskMt$ \\ \hline \hline
    \end{tabular}
    \caption{{ The ratio $M^{*}_{\rm d}/M^{*}_{\rm s}$ and $\rho_{\rm d}/\rho_{\rm s}$ values for various combinations of dust mass functions derived in this work and stellar mass functions for different populations from this work and from M16a and M16b$^{16}$.} }
    \label{tab:Ratio}
\end{table}

We note that the uncertainties on the datapoints in the GSMFs quoted in M16a and M16b are based on an on-sky jackknife analysis. As we described in Section~\ref{DMFshape}, this estimate will include a cosmic variance uncertainty component. Since we cannot disentangle the inherent cosmic variance from their values we choose instead to use the same percentage uncertainty in the integrated stellar mass density as our percentage uncertainty in the integrated dust mass density. Since the errors in dust mass are larger than for stellar mass for our dataset, the uncertainty in the integrated stellar mass density should not be larger than for our integrated dust mass density for the same sample; as such this estimate of the error is a conservative one.

\section{Conclusions}
\label{sec:conc}

We measure the DMF for galaxies at $0.002 \le z \le 0.1$ using a modified $V_{\rm max}$ method for 15,750 sources.  This represents the largest study of its kind both in terms of numbers of galaxies and the volume probed at this redshift range.  Dust masses are derived using the {\sc magphys} SED fitting tool given photometric fluxes from {\sc \lambdar}. { Despite the sources that have measurements below $3\,\sigma$ in one or more {\it Herschel} SPIRE bands, we show that the {\sc \magphys} derived errors of 0.4, 0.18, 0.14 and 0.1\,dex for galaxies with flux $>3\,\sigma$ in zero, one, two and three SPIRE bands do reasonably represent the uncertainty in dust mass. }

We use a { single} Schechter function fit to our data to compare with previous observations and we extend
the DMF down to lower dust masses than probed before, constraining the faint end slope below $10^{5}\, M_{\odot}$.  Our main findings are:

\begin{itemize}

\item We compare the dust mass function derived using the traditional $pV_{\rm max}$ method and the BBD method which allows us to incorporate selection effects, and find both ultimately produce similar results. The best fitting single Schechter function for the $z\le 0.1$ DMF has $\alpha = \pVmaxalpha$, $M^* = (\pVmaxMstar)
  \times 10^7\,h^2_{70}\, M_{\odot}$, $\phi_* = (\pVmaxphistar)
  \times 10^{-3}\,h^3_{70}\,\rm Mpc^{-3}\,\rm dex^{-1}$, and
  $\Omega_{\rm d} = (\pVmaxOmegad) \times 10^{-6}$. There is an additional uncertainty from cosmic variance of 7-17\%.


  \item { We find that a double Schechter function is formally a better fit to the observed DMF, though including cosmic variance would make the improvement in the fit not significant. Also given the variation of errors as a function of mass and the lack of correspondence to physical subsets of galaxies, we do not have confidence that the double Schechter function represents a better description of the DMF. }
  

\item We find that there is a discrepancy between the observed and
  predicted dust mass functions derived from the semi-analytic models of \citet{Popping2017}. This is largest at the high-stellar mass end, with
  the model predictions of $M^*$ higher by $0.5\,$dex compared to our observed,
  Schechter functions. The likely cause
  for the discrepancy is that the Popping model uses a relationship
  between dust and stellar mass which is inconsistent with properties observed in
  local galaxies samples such as the {\it Herschel}-ATLAS, the HRS and the DGS, and also with our sample of GAMA sources: the models produce high
  stellar mass galaxies with dust masses far higher than is
  observed.  This discreprancy is alleviated somewhat when we compare with the predicted DMF from cosmological hydrodynamical simulations \citep{McKinnon2017} who use longer grain growth timescales. This reduces the amount of dust formed in high mass galaxies; however, \cite{McKinnon2017} under-predict the number of high dust mass galaxies compared to our observations, although the limited volume of their simulation does not allow a proper comparison. Both sets of theoretical predictions also fail to match the observed volume density of low dust mass galaxies.
  Our dataset thus provides a useful benchmark for models.

\item Splitting our sample into early and late-type on the basis of
  morphology and colour (to a redshift limit of $z\le 0.06$), results in DMFs with very different
  shapes. The late-type DMF is well represented by a Schechter function,
  whereas the ETG DMF is not. The LTG DMF has far higher space
  density at a given dust mass.  We derive dust mass densities of $\Omega_{\rm d} = (\MoffnotElowOmegadinf) \times 10^{-6}$ and  $\Omega_{\rm d} = (\MoffElowOmegadinf) \times 10^{-6}$ for late types and early types respectively. In total there is $\sim 10$ times more dust
  mass density in late-type galaxies compared to early-types at $ 0.002 < z \le 0.06$.

\item In comparing our DMF to the GSMFs from W17 and
  \cite{Moffett2016_gals,Moffett2016_disks}, it is possible to scale from the { LTG galaxy stellar mass function (GSMF) to the LTG DMF using a ratio of $\rho_{\rm d}/\rho_{\rm s} = (\rhoLTGBnLTGMt)\times 10^{-4}$. Similarly, we show that one can scale from the disk GSMF to the LTG DMF using the ratio $\rho_{\rm d}/\rho_{\rm s} = (\rhoLTGBndskMt)\times 10^{-4}$. } We caution that using Schechter values derived from Schechter function fits to the ETG DMF and Elliptical GSMF may be inadvisable since neither are well-fitted by a Schechter function, although scaling from the Elliptical GSMF to the ETG DMF by multiplying by $\rho_{\rm d}/\rho_{\rm s} = (\rhoETGBnEMt)\times 10^{-4}$ returns a reasonable representation of the ETG DMF around the knee.

\end{itemize}

\section*{Acknowledgements}

We acknowledge our anonymous referee for their helpful input. We acknowledge Edo Ibar and Michal Micha{\l}owski for their comments on earlier drafts of this paper and acknowledge Ryan McKinnon for sharing his dust mass function for this paper. RAB, HLG, SJM, and LD acknowledge support from the European Research Council (ERC) in the form of Consolidator Grant {\sc CosmicDust} (ERC-2014-CoG-647939, PI H\,L\,Gomez). LD, SJM and NB acknowledge support from European Research Council Advanced Investigator Grant COSMICISM, 321302. CJRC acknowledges support from the ERC in the form of the 7$^{\rm th}$ framework programme (FP7) grant DustPedia (PI Jon Davies, proposal 606824). GAMA is a joint European-Australasian project based around a spectroscopic campaign using the Anglo-Australian Telescope. The GAMA input catalogue is based on data taken from the Sloan Digital Sky Survey and the UKIRT Infrared Deep Sky Survey. Complementary imaging of the GAMA regions is being obtained by a number of independent survey programmes including GALEX MIS, VST KiDS, VISTA VIKING, WISE, Herschel-ATLAS, GMRT and ASKAP providing UV to radio coverage. GAMA is funded by the STFC (UK), the ARC (Australia), the AAO, and the participating institutions. The GAMA website is \url{http://www.gama-survey.org/}. The Herschel-ATLAS is a project with Herschel, which is an ESA space observatory with science instruments provided by European-led Principal Investigator consortia and with important participation from NASA. The {\it H-}ATLAS website is {\url{http://www.H-ATLAS.org/}}.  This research made use of Ned Wright's cosmology calculator {\url{http://www.astro.ucla.edu/~wright/CosmoCalc.html}} \citep{Wright2006}. Fig.~2 was made with the python seaborn package {\url{ http://seaborn.pydata.org/index.html}}.




\bibliographystyle{mnras}
\bibliography{DustMassFunction.bib} 




\appendix

\section{FIR Constraints on Dust Temperature}
\label{Appendix:Temperature}

\begin{figure*}
\includegraphics[trim=10mm 4mm 10mm 5mm clip=true,width=0.95\textwidth]{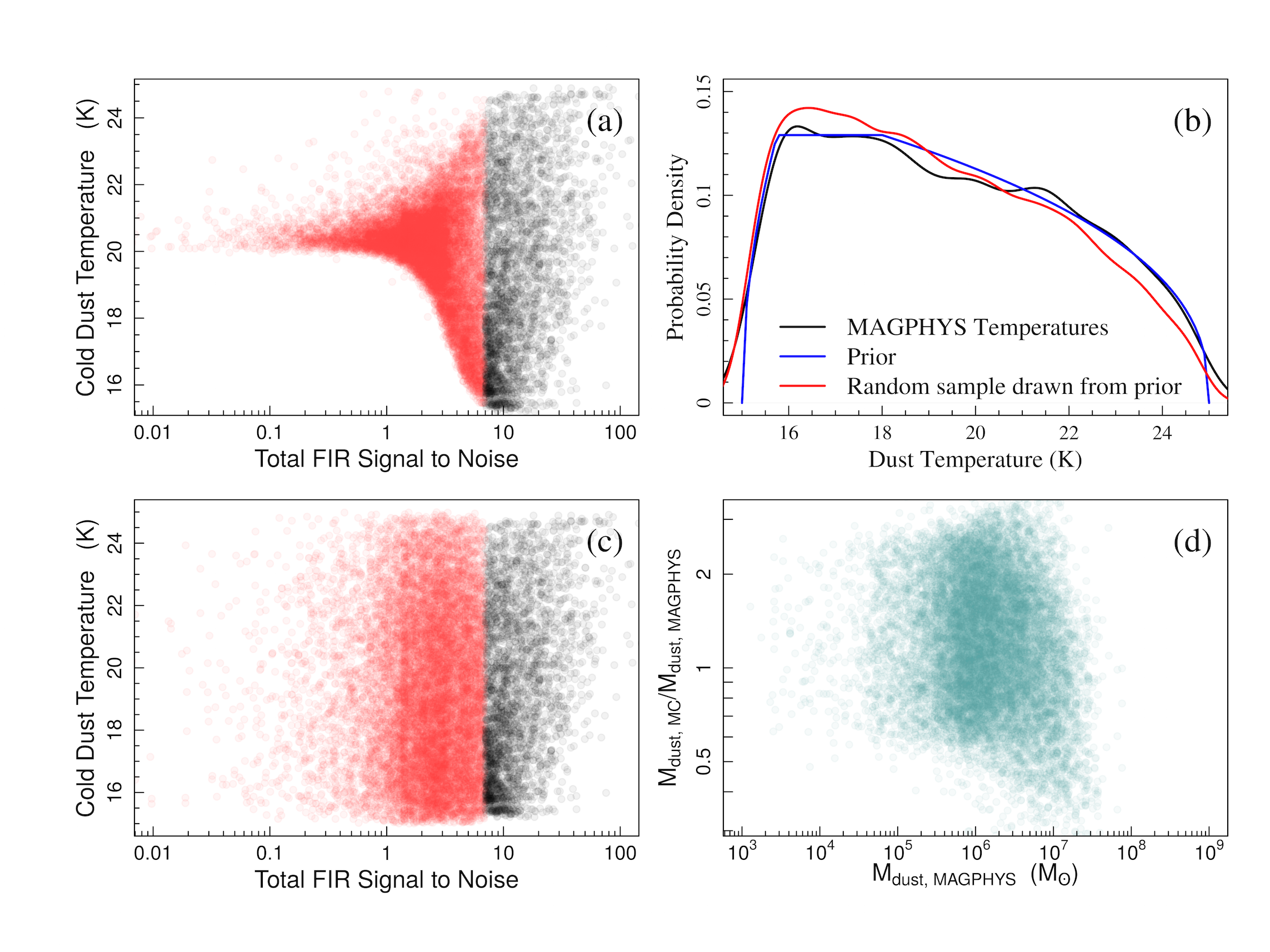}
 \caption{{ (a)} The cold dust temperature from \magphys~as a
   function of total FIR signal to noise, in red all those sources
   below 7$\,\sigma$, in black all those sources with 7$\,\sigma$ or
   greater total FIR flux. { (b)} In black the probability density
   function (PDF) of temperatures from \magphys~for those galaxies
   with a total FIR signal to noise greater than 7 using kernel
   density estimation (KDE), in blue the prior we use to describe this
   PDF, in red a KDE of one example of a random draw of temperatures
   from this prior.  { (c)} In red we show one MC simulation of new
   cold dust temperatures for sources below 7$\,\sigma$ total FIR
   flux, in black the \magphys~cold dust temperatures for sources with
   7$\,\sigma$ or greater total FIR flux. { (d)} One MC realisation
   of the ratio of dust masses adjusted for their MC temperature as a
   function of their dust mass as assigned by \magphys.  }
 \label{fig:temp4panel}
\end{figure*}

\begin{figure}
 \includegraphics[width=\columnwidth]{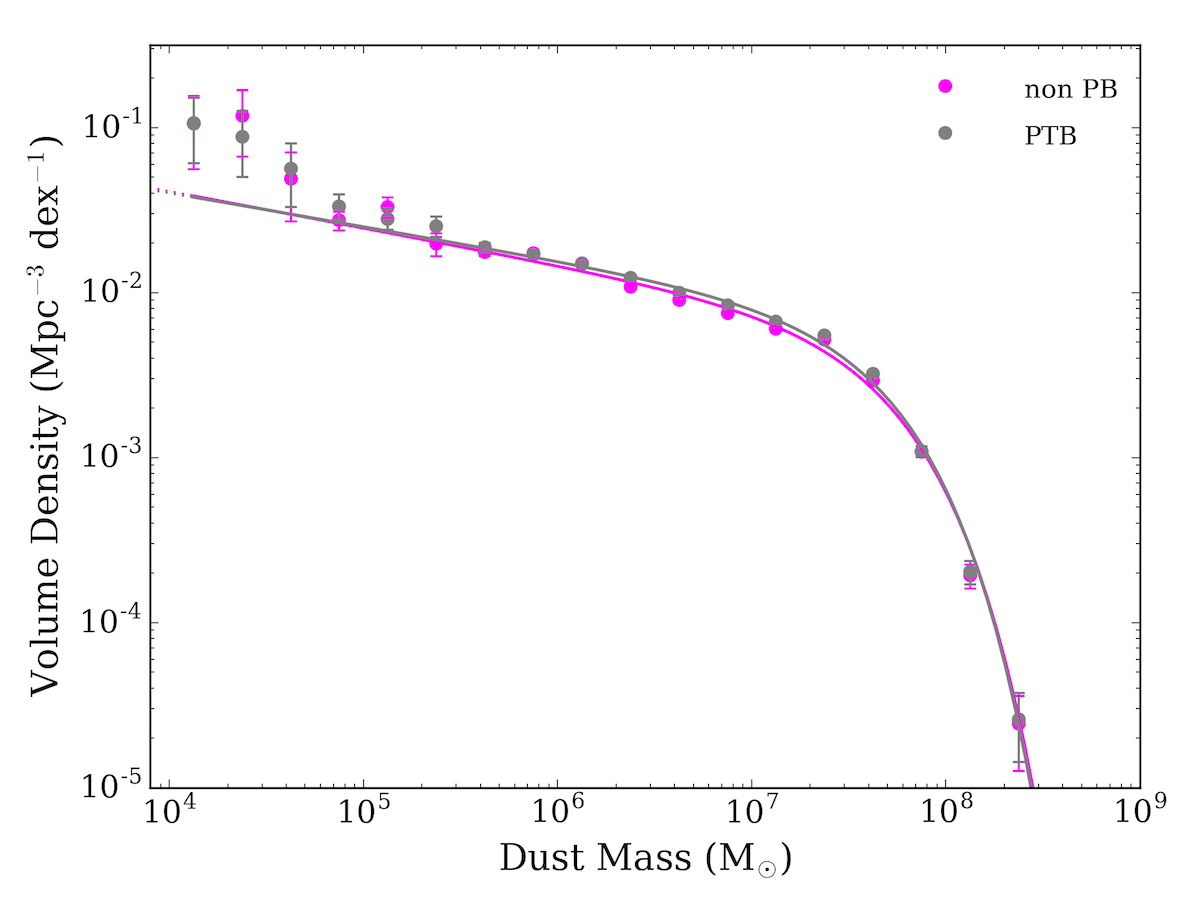}
 \caption{ The $pV_{\rm max}$ dust mass functions for the GAMA/{\it H-}ATLAS sources, firstly showing the SB DMF as seen in Section \ref{sec:dmf} in magenta, and secondly the DMF resulting from the perturbed temperature bootstrap (PTB) method shown in grey. The data points show the observed values and the solid lines are the best fitting ($\chi^2$) single Schechter functions to the data for their respective fitted regions, beyond this we show extrapolations down to $10^{4} M_{\odot}$ as dashed lines. Error bars are derived from a bootstrap analysis and the data points have been corrected for over and under densities in the GAMA fields (see W17). }
 \label{fig:TempMC_DMF}
\end{figure}

{

  In section 3, we found that when a galaxy has poor FIR constraints,
  the cold dust temperature PDF from \magphys\ simply reflects the
  prior temperature distribution. Since the FIR data provide the only
  constraints on the cold temperature, this is the correct result that
  \magphys\ should output. However, taking the median of the PDF in these
  cases may lead to a systematic bias in the temperature, which would
  in turn lead to a bias the dust mass.  Roughly one third of our
  sample has a combined FIR signal to noise of less than 1, so
  potentially, it is a significant effect. Here we attempt to quantify
  how this effect propagates into in our estimates of the DMF. In
  panel (a) of Fig. A1 we show the median cold dust temperature output
  by \magphys\ as a function of total FIR signal to noise. As the
  total FIR signal to noise decreases there is a clear trend for
  galaxies to be assigned a cold dust temperature nearer to 20\,K, the
  median of the \magphys\ temperature prior.  Panel (b) of Fig. A1
  shows the distribution of cold dust temperatures for sources with a
  total FIR signal to noise of at least $7\sigma$, where the
  constraints on the dust temperature are very good. Assuming the
  underlying temperature distribution is independent of observed
  signal to noise, we can use this as a new prior for the temperatures
  of sources which have poor FIR constraints.

The analysis we perform is another form of perturbed bootstrap, where,
instead of perturbing the masses based on the 16 and 84 percentiles
output by \magphys, we perturb their cold dust temperatures to match
the new prior temperature distribution.  As in the PB method we first
resample from the underlying dataset to give each bootstrap
realisation. We keep the temperatures of the high signal to noise
sources the same, but for all other galaxies we assign new
temperatures using the prior from the high signal to noise
sample. This means that the resulting temperature distribution for all
sources matches the prior. This is shown for one realisation in panel
(c) of Fig. A1.  Finally we adjust the galaxy dust masses in proportion
to the change in black body luminosity at 250\,$\mu$m given the
original and re-assigned temperatures. We refer to this method as the
perturbed temperature bootstrap (PTB). The ratio of the PTB and
\magphys\ temperatures as a function of dust mass from \magphys\ for
one realisation can be seen in panel (d) of Fig. A1. The figure shows
that the dust mass can increase or decrease as a result of the
temperature MC, but generally the dust masses increase as a result of
the new temperature.

Fig A2 and Table A1 show the resulting DMF and best fit Schechter
function parameters along with our simple bootstrap (SB) results. The differences are
very small, showing that the poorly constrained temperatures do not
introduce a significant bias in our DMF estimates, even though they
make up a significant fraction of our sample.  The derived $\alpha$
and $M_{\rm s}$ values both agree within uncertainties. The value of
$\phi_*$ is 2\,$\sigma$ or $\sim 16\%$ higher for the PTB DMF.  Overall the
revised temperatures lead to a difference in the cosmic dust density
of only $\sim 6\%$. Thus we are confident that any bias from poor mass
constraints for the lower signal to noise sources in our sample is at
the level of a few percent.

\begin{table}
\centering
\begin{tabular}{ccc}
\hline \hline
                                    & SB                & PTB                        \\ \hline
\multirow{2}{*}{$\alpha$}           & \multirow{2}{*}{$\pVmaxalpha$} & \multirow{2}{*}{$\pVmaxTmpalpha$} \\
                                    &                                   &                                   \\
$M^{*}$                               & \multirow{2}{*}{$\pVmaxMstar$}  & \multirow{2}{*}{$\pVmaxTmpMstar$}  \\
($10^7\,M_{\odot}$)             &                                   &                                   \\
$\phi^{*}$                            & \multirow{2}{*}{$\pVmaxphistar$}  & \multirow{2}{*}{$\pVmaxTmpphistar$}  \\
($10^{-3}\,h^3_{70}\,\rm Mpc^{-3}\,dex^{-1}$) &                                   &                                   \\
$\Omega_{\rm d}$                          & \multirow{2}{*}{$\pVmaxOmegad$}  & \multirow{2}{*}{$\pVmaxTmpOmegad$}  \\
($10^{-6}$)            &                                   &

 \\
\hline \hline

\end{tabular}
    \caption{Schechter function fit values for dust mass functions resulting from the SB and PTB DMF analysis. The fits in this work include the density-weighted corrections from W17.}
    \label{tab:appendixSchechterTabTemps}
\end{table}

}
\section{Changing the bivariate dust mass function to stellar mass-surface brightness}
\label{Appendix:axes}

\begin{figure*}
\begin{center}
 \includegraphics[trim=0mm 8mm 9mm 5mm clip=true,width=0.8\textwidth]{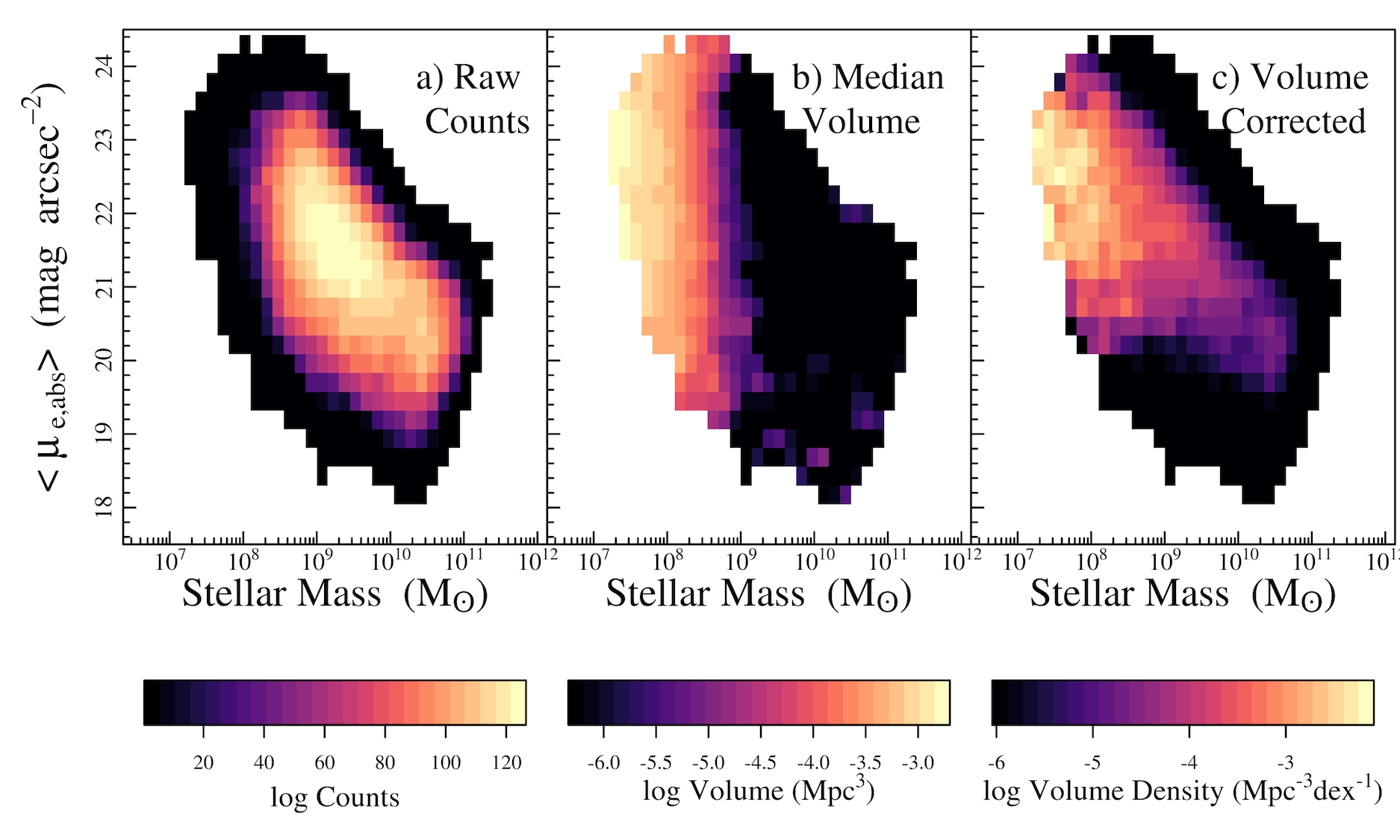}
 \caption{The BBD for our sample with $r$-band magnitude and surface brightness as the two `axes' (W17)  a) Raw counts in stellar mass/surface brightness bins, b) Median volume in the stellar mass/surface brightness bins, c) Weighted counts, i.e. volume density in the stellar mass/surface brightness bins. { Each of the panels represents the BBD resulting from the average of 1000 Monte Carlo simulations where we perturb the stellar mass and surface brightness within their associated uncertainties}.}
 \label{fig:SB_bbd}
\end{center}
\end{figure*}

\begin{figure*}
  \centering
  \includegraphics[trim=10mm 5mm 0mm 0mm clip=true,width=0.48\textwidth]{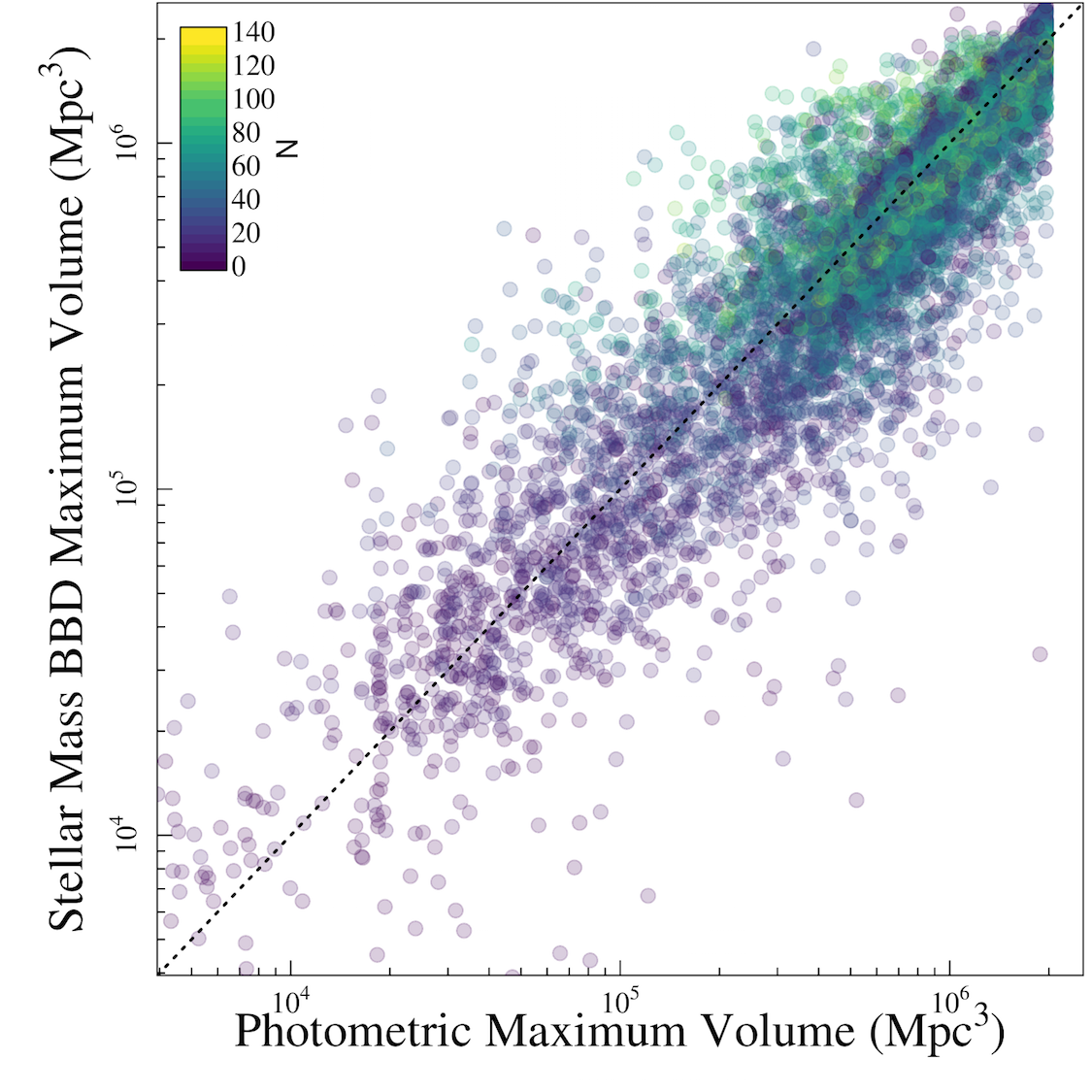}
 \includegraphics[trim=0mm 6mm 8mm 2mm clip=true,width=0.5\textwidth]{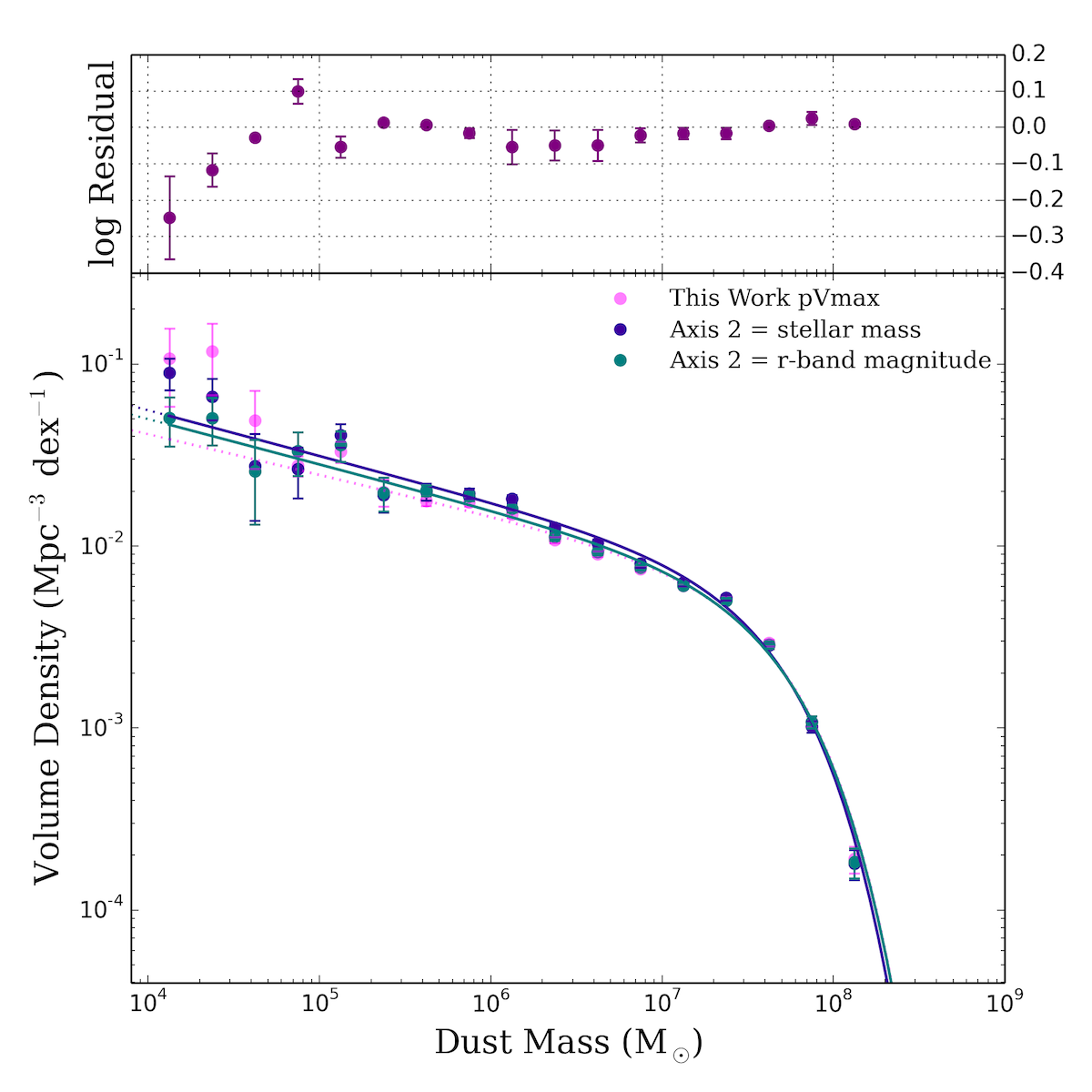}
 \caption{{\it Left:} The maximum effective volumes for our galaxies at $z<0.1$ derived using the $pV_{\rm max}$ method (x-axis), and BBD method using stellar mass and surface brightness as the two selection features (y-axis). The colour of the points is determined by the number of galaxies in the BBD bin that each galaxy resides in (Fig \ref{fig:SB_bbd}), as shown by the colour bar in the top left corner.  {\it Right:} The DMF derived using a stellar mass and surface brightness BBD (blue) and an $r$-band magnitude and surface brightness BBD (green) for the GAMA/{\it H-}ATLAS sources at $z\le 0.1$, also shown are the $pV_{\rm max}$ values for comparison in magenta.  The data points show the observed values and the solid lines are the best-fitting ($\chi^2$) single Schechter functions to the data. Error bars are derived from a bootstrap analysis and the data points have been corrected due to over and under densities in the GAMA fields (see W17). The residual between the two BBD DMFs is shown in the top panel, as points with error bars in purple.}
 \label{fig:appendixdmf}
\end{figure*}

We also performed the BBD analysis described in Section~\ref{sec:dmf_volEst} with stellar mass and surface brightness as the two `axes'. We show the raw counts of galaxies in $M_{\rm s}$/$\mu_{\rm e,abs}$ bins in Fig.~\ref{fig:SB_bbd}a, the median volume in Fig.~\ref{fig:SB_bbd}b, and the weighted counts (volume density) in Fig.~\ref{fig:SB_bbd}c. In comparison to Fig.~\ref{fig:DM_bbd}, we can see that the shapes and trends of the BBDs with $r$-band$/\mu_{\rm e,abs}$ or $M_{\rm s}$/$\mu_{\rm e,abs}$ as the second axis are unsurprisingly very similar. The main difference between the two is that the shape of the latter reflects a slightly stronger evolution of stellar mass compared to $r$-band magnitude with surface brightness.

Fig.~\ref{fig:appendixdmf} reproduces Fig.~\ref{fig:BBDVmaxComp} but now with stellar mass as the second `axis' of the BBD.   The $pV_{\rm max}$ and $M_{\rm s}$/$\mu_{\rm e,abs}$ are coloured by the number of galaxies in the BBD bin containing that galaxy. The largest deviations from the 1:1 line are seen when the galaxy lies in a more sparsely populated bin, generally these volumes are low and are therefore subject to large cosmic variance. The $pV_{\rm max}$ values are systematically higher by 1.2\% on average than those derived from the $M_{\rm s}$/$\mu_{\rm e,abs}$ BBD method, which gives an average offset of 6\% in the binned DMF values.  This is because the largest differences in the binned DMF values come from the low dust mass end where BBD bins are more likely to be below the required 4 galaxies per bin. The scatter about the 1:1 line is large compared to Fig.~\ref{fig:BBDVmaxComp} since a galaxy's $r$-band magnitude has more reason to impact the sample selection than its stellar mass, and so ultimately we decide to show the $r$-band magnitude BBD in the main body of the paper.

Finally, Fig.~\ref{fig:appendixdmf} compares the resulting DMFs from using $r$-band magnitude or stellar mass as the second `axis' of the BBD, where the surface brightness is used for the remaining axis. The Schechter fit parameters are compared in Table~\ref{tab:appendixSchechterTab}, and the residuals between the datapoints resulting from the both the $M_{\rm s}$/$\mu_{\rm e,abs}$ and $r$-band magnitude/$\mu_{\rm e,abs}$ BBD are shown in Fig.~\ref{fig:appendixdmf}. The Schechter fit parameters for both sets of BBD DMFs agree within uncertainties; however, the $r$-band magnitude BBD is closer to the $pV_{\rm max}$ fit parameters. The $pV_{\rm max}$ DMF points are offset from the stellar mass BBD by $\sim$6\,per\,cent, and the $r$-band magnitude BBD by $\sim$3\,per\,cent on average.

We also tried using dust mass and stellar mass as the two axes for the BBD since there is a dichotomy seen for ETGs and LTGs as well as for sources with and without a $3\,\sigma$ measurement in any SPIRE band. There was no significant departure from either the $r$-band magnitude or the stellar mass BBD and so we cannot say that either split strongly affects the resulting BBD $V_{\rm max}$ since it is not possible to deconvolve the effects that each split may have on the accessible volumes.

\begin{table}
\centering
\begin{tabular}{ccc}
\hline \hline
\multirow{2}{*}{}                   & \multicolumn{2}{c}{Axis 2}                                            \\
                                    & Stellar Mass                & $r$-band Magnitude                        \\ \hline
\multirow{2}{*}{$\alpha$}           & \multirow{2}{*}{$\bbdSBalpha$} & \multirow{2}{*}{$\bbdDMalpha$} \\
                                    &                                   &                                   \\
$M^{*}$                               & \multirow{2}{*}{$\bbdSBMstar$}  & \multirow{2}{*}{$\bbdDMMstar$}  \\
($10^7\,M_{\odot}$)             &                                   &                                   \\
$\phi^{*}$                            & \multirow{2}{*}{$\bbdSBphistar$}  & \multirow{2}{*}{$\bbdDMphistar$}  \\
($10^{-3}\,h^3_{70}\,\rm Mpc^{-3}\,dex^{-1}$) &                                   &                                   \\
$\Omega_{\rm d}$                          & \multirow{2}{*}{$\bbdSBOmegad$}  & \multirow{2}{*}{$\bbdDMOmegad$}  \\
($10^{-6}$)            &                                   &

 \\
\hline \hline

\end{tabular}
    \caption{Schechter function fit values for dust mass functions resulting from the BBD with the second axis being stellar mass, and for the second axis being $r$-band magnitude, both have surface brightness on the first axis. The fits in this work include the density-weighted corrections from W17.}
    \label{tab:appendixSchechterTab}
\end{table}

\section{Changing the minimum mass limit used in fitting the Dust Mass Function}
\label{Appendix:mmin}

\begin{figure*}
\includegraphics[trim=10mm 4mm 10mm 5mm clip=true,width=0.95\textwidth]{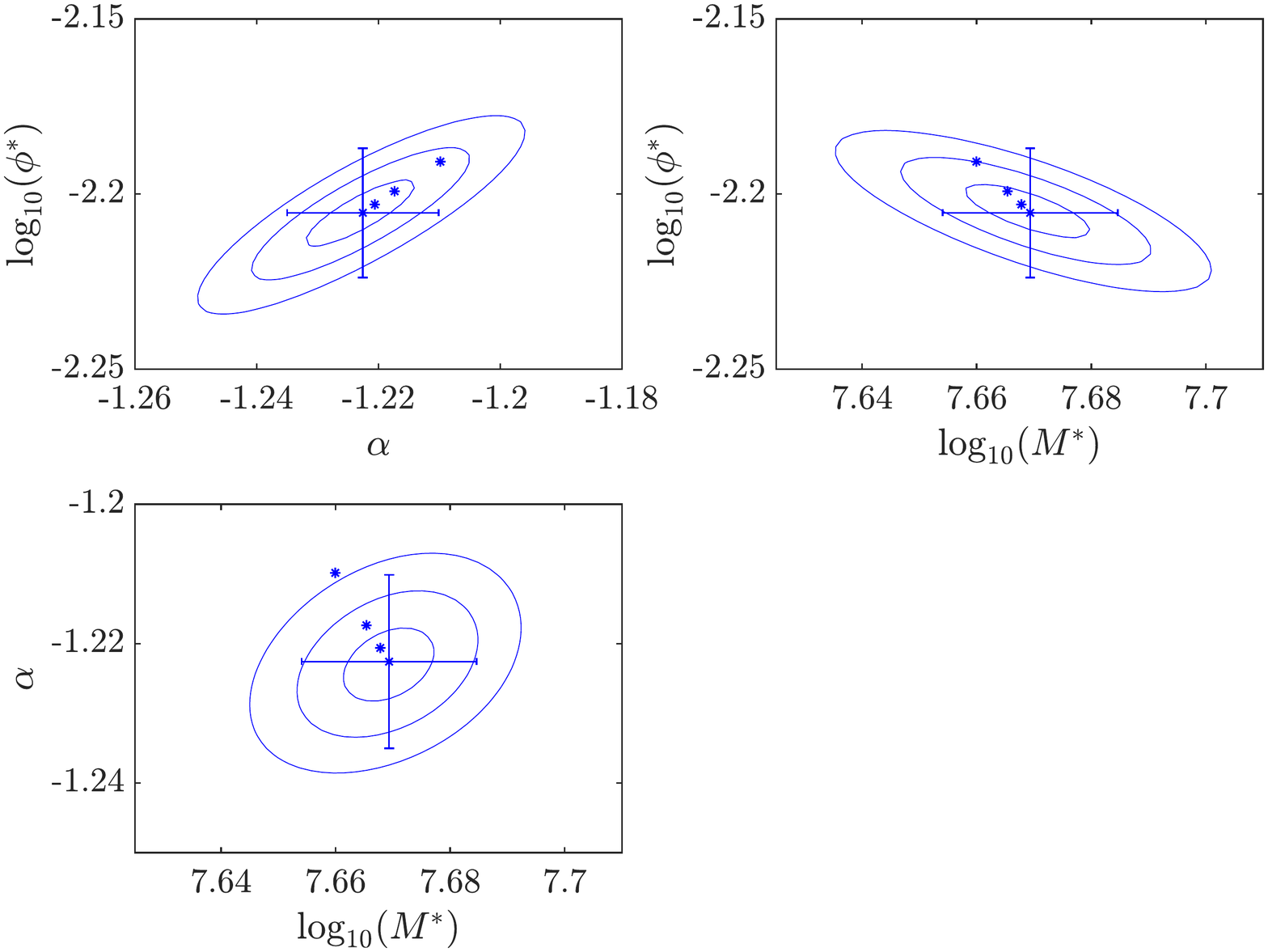}
 \caption{ { The confidence intervals for the $pV_{\rm max}$ Schechter fits (SF) to the single dust mass function derived in this work (blue ellipses) as a function of minimum mass  chosen for the fit,  $\log_{10}(M_{\rm min}/M_{\odot}) = 4, 4.5, 5$ and $5.5$. The contours and error bars are centred on the fit with $\log_{10}(M_{\rm min}/M_{\odot}) = 4$. Error bars on the fit parameters are taken from the $\Delta \chi^2=1$ for each parameter.  Note that $\phi^*$ is in units of $\rm Mpc^{-3}\, {\rm dex}^{-1}$. } }
 \label{fig:mmin_conf}
\end{figure*}

{
Fig.~\ref{fig:mmin_conf} compares the resulting  single Schechter function fit parameters derived using different low mass limits ranging from $10^{4}$ to $10^{5.5}M_{\odot} $. We see a convergence of the derived $\alpha$, $M^*$ and $\phi^*$ when using the single fit with $M_{\rm min}<10^{4.5}M_{\odot}$. Beyond this point we see that there is a strong dependence on the best-fitting parameters with $M_{\rm min}$. 
}

\bsp	
\label{lastpage}
\end{document}